\documentclass[10pt,a4paper,aps,showpacs,showkeys,
				amsmath,amssymb,pra,floatfix,
				twocolumn,
				superscriptaddress,
				longbibliography
				]{revtex4-1}

\usepackage[mathscr]{euscript}
\usepackage[normalem]{ulem}

\usepackage{microtype}

\usepackage{hyperref}

\usepackage{amsmath}
\usepackage{amssymb}
\usepackage{slashed}
\usepackage[usenames,dvipsnames]{xcolor}
\usepackage{graphicx}
\usepackage{grffile}
\usepackage[squaren]{SIunits}
\usepackage{soul}
\renewcommand{\mathbf}{\boldsymbol}

\usepackage{cleveref}
\hypersetup{colorlinks=true,citecolor=blue}

\DeclareSymbolFont{rsfs}{U}{rsfs}{m}{n}
\DeclareSymbolFontAlphabet{\mathrsfs}{rsfs}

\renewcommand{\d}{\mathrm d}
\newcommand{\floating}{{moving }}
\newcommand{\Stokes}{{polarization }}
\let\ud\d
\let\vec\boldsymbol
\let\up\uparrow
\let\down\downarrow

\newcommand{\tr}{{\rm tr}\,}
\newcommand{\TR}[1]{\, \tr \! \left[#1\right] }
\newcommand{\Ai}{\mathrm{Ai}\,}

\begin{document}

\title{Theory of Radiative Electron Polarization in Strong Laser Fields}
\author{D.~Seipt}
\email{d.seipt@lancaster.ac.uk}

%\email{dseipt@umich.edu}

\affiliation{Physics Department, Lancaster University, Bailrigg, Lancaster LA1 4YW, United Kingdom}
\affiliation{The Cockcroft Institute, Daresbury Laboratory, WA4 4AD, United Kingdom}
%\affiliation{Center for Ultrafast Optical Science, University of Michigan, Ann Arbor, Michigan 48109, USA}

\author{D.~Del Sorbo}
\affiliation{York Plasma Institute, Physics Department, University of York, York YO10 5DQ, United Kingdom}

\author{C.~P.~Ridgers}
\affiliation{York Plasma Institute, Physics Department, University of York, York YO10 5DQ, United Kingdom}

\author{A.~G.~R.~Thomas}
\affiliation{Center for Ultrafast Optical Science, University of Michigan, Ann Arbor, Michigan 48109, USA}

\date{\today}

	\begin{abstract}
Radiative polarization of electrons and positrons through the Sokolov-Ternov effect is important for applications in high-energy physics. Radiative spin-polarization is a manifestation of quantum radiation reaction affecting the spin-dynamics of electrons. We recently proposed that an analogue of the Sokolov-Ternov effect could occur in the strong electromagnetic fields of ultra-high-intensity lasers, which would result in a  build-up of spin-polarization in femtoseconds. In this paper we develop a density matrix formalism for describing beam polarization in strong electromagnetic fields. We start by using the density matrix formalism to study spin-flips in non-linear Compton scattering and its dependence on the initial polarization state of the electrons. Numerical calculations show
    a radial polarization of the scattered electron beam in a  circularly polarized laser, and we
	find azimuthal asymmetries in the polarization patterns for ultra-short laser pulses. A degree of polarization approaching $\unit{9}{\%}$ is achieved after emitting just a single photon. We develop the theory by deriving a local constant crossed field approximation (LCFA) for the polarization density matrix, which is a generalization of the well known LCFA scattering rates. We find spin-dependent expressions that may be included in electromagnetic charged-particle simulation codes, such as particle-in-cell plasma simulation codes, using Monte-Carlo modules. In particular, these expressions include the spin-flip rates
	for arbitrary initial polarization of the electrons. The validity of the LCFA is confirmed by explicit comparison with an exact QED calculation of  electron polarization in an ultrashort laser pulse.
	\end{abstract}

	\maketitle

	\section{Introduction}
    
	Electrons are fermions with an intrinsic spin-angular momentum of $\hbar/2$, with an associated intrinsic magnetic (dipole) moment
	of $\mu = - g_e e \hbar / 4 m$ with charge $e$, mass $m$, and with $g_e$ being the gyromagnetic ratio \cite{RevPartPhys}.
	The electron spin polarization can be manipulated by using electromagnetic fields \cite{book:Jackson}.

	The scattering of spin-polarized particles is an important aspect of
    high-energy physics.
	For instance, the polarized deep inelastic scattering of polarized leptons on
	polarized protons revealed intriguing details on the spin-structure of the constituents of the proton
	\cite{Anselmino:PhyRep1995,Bass:ModPhysLettA2009}, and polarized beams
	have been used in investigations of parity non-conservation effects \cite{Prescott:PLB1978,Labzowsky:PRA2001}.
	Moreover, using polarized beams in upcoming high-energy lepton-lepton colliders can help
	suppress the standard-model background in searches for new physics beyond the standard model \cite{Vauth:JMPConfSer2016}.

	In conventional accelerators the spin-dynamics is well understood \cite{Mane:RPP2005},
    and very high degrees of polarization $>\unit{80}{\%}$
	can be achieved routinely by injecting a polarized beam and maintaining the polarization throughout the acceleration process
	\cite{Subashiev:LowDim1999,Mamaev:APL2008}. Even if the beams are initially unpolarized, in a storage ring they will
	spin-polarize eventually due to the emission of synchrotron radiation, the so-called Sokolov-Ternov-effect.
	This radiative spin-polarization can lead to an equilibrium polarization degree of
	up to $\unit{92.4}{\%}$, with the electrons' spins anti-parallel to the magnetic field \cite{book:Sokolov,Ternov:PhysUspekh1995}.
	The process is rather slow, on the order of minutes or hours, since the relevant field strengths are quite low.

	In novel accelerator concepts that employ laser driven plasma wake field structures the generated electron beams are usually
	unpolarized \cite{Esarey:RevModPhys2009}, and 
	in most studies of in high-intensity laser-plasma interactions the electron polarization is not taken into account.
	Vieira et al. \cite{Vieira:PRSTAB2011} investigated the spin precession of polarized electrons that are injected into
	a laser wakefield accelerator to determine how strongly the beam depolarizes in the acceleration process.

	The question that arises is: Can the interaction of an unpolarized electron beam with an intense laser pulse be used to
	polarize the electrons? The laser-driven electrons will emit gamma ray photons due to the non-linear Compton process.
	It has been shown recently in Ref.~\cite{DelSorbo:PRA2017,DelSorbo:PPCF2018} that electrons orbiting and radiating in the
	magnetic anti-node of a standing
	intense laser wave do indeed spin-polarize within a few femtoseconds for laser intensities exceeding
	$ \unit{5 \times 10^{22}}{\watt\per\centi\metre^2}$.
	In order to generalize these results we investigate in this paper in detail how the non-linear Compton scattering
	probabilities depend on the spin polarization of the electrons.

	Recent experiments gave strong evidence for the quantum nature of radiation reaction \cite{Cole:PRX2018,Poder:PRX2018}
	in collisions of electron beams with high-intensity laser pulses, i.e. on
	the back-reaction of the radiation emission on the emitting electron orbital motion, the dominant process of which is non-linear Compton scattering.
	In this regard, radiative spin-polarization can be seen as another aspect of quantum radiation reaction, affecting the spin dynamics.
	Ultra-fast spin-polarization could possibly be studied in ultra-intense laser electron beam collisions with present day or near future laser facilities.

	In all experiments on radiative polarization so far the polarization times are very long
	because spin-flip transitions are suppressed by a small
	quantum efficiency parameter $\chi_e =  e \sqrt{ | (F^{\mu\nu} p_\nu)^2 |}/m^3  \ll 1$ \cite{book:Sokolov}.
	In a strong laser field, $\chi_e  = \xi b \sim 1$, where
	$\xi = 8.55 \sqrt{\lambda^2[\micro\metre ] I [10^{20}\watt/\centi\metre^2 ]}$ is the classical nonlinearity parameter
	with laser wavelength $\lambda$ and intensity $I$.
	For a strong laser field $\xi \gg 1$ and $\chi_e\sim1$ the spin can rapidly polarize on femtosecond time scales
	\cite{DelSorbo:PRA2017,DelSorbo:PPCF2018}.
	$b= k.p/m^2$ is the quantum energy parameter, i.e.~the laser frequency in the electron rest frame.
	In perturbative QED ($\xi \ll 1 $) spin effects in Compton scattering
	become important only in the quantum regime, when $b$ is not small.
	In fact, upon re-inserting $\hbar$ the parameter $b = \mathcal O(\hbar)$, confirming that spin is a quantum property,
	and spin-flip transitions are not occurring in the classical low-energy limit $b\to0$.

	In calculations of strong-field QED processes the spin degree of freedom is automatically accounted for
	when working with solutions of the Dirac equation.
	In fact, significant differences have been found when comparing the scattering of
    unpolarized spin 1/2 particles with calculations in scalar QED
    for non-linear Compton scattering, pair production, or channelling crystals fields
	\cite{Kirsebom:PRL2001,Panek:PRA2002b,Dumlu:PRD2011,Boca:NIMB2012,Jansen:PRD2016}. 
	Most studies of spinor QED refer to unpolarized electrons where the electron spin is averaged over.
	More detailed calculations of the non-linear Compton scattering probabilities with regard to
	the spin and spin polarization have been performed for monochromatic laser fields 
	\cite{Bagrov:INCB1989,Bolshedvorsky:PTP2000,Bolshedvorsky:RusPhysJ2000,
		Ivanov:EPJC2004,Karlovets:PRA2011},
	short laser pulses \cite{Boca:NIMB2012,Krajewska:LasPB2013,Krajewska:PRA2014c},
	or constant crossed fields \cite{King:PRA2013,King:PRA2015}. Moreover, in extremely strong fields
    electrons may polarize without radiating via electromagnetic self-interaction \cite{Meuren:PRL2011}.
	We mention also that the spin-dependence of other strong-field processes has been discussed recently in the literature
	\cite{Ahrens:PRL2012,Barth:PRA2013b,Klaiber:PRA2014,Zille:JPB2017,Zille:PRA2017,Dellweg:PRA2017}.

	In this paper we give a precise and thorough description of the polarization properties of
	the final state electrons in non-linear Compton scattering.
	Using the density matrix formalism for interaction of electrons with short intense laser pulses of arbitrary duration, shape, and polarization
	we derive compact expressions for the properties of the scattered electrons.
	We are interested not only in the spin-dependent scattering probabilities, but specifically on how
	the degree of polarization of the electrons changes during the non-linear Compton process.
	Numerical calculations of the electron polarization of the final electrons are presented for collisions of high-energy electron with an
	ultra-short intense laser pulse.
	We derive analytically the locally constant crossed field approximation (LCFA) expressions for the final electron polarization 		
	density matrix as a generalization of the spin-dependent LCFA scattering rates.
	Using these results we find that electrons can spin-polarize also in asymmetric ultra-short linearly polarized laser pulses.
	The LCFA polarization density matrix can serve as the basis for a spin-dependent full-scale QED laser-plasma simulation code,
	in analogy to the LCFA emission rates used in QED-PIC codes.
	To validate our analytical results we compare them with an exact QED calculation of the degree of polarization in an ultrashort laser pulse 
	and find excellent agreement.

	Our paper is organized as follows:  
	In Section \ref{sect:theory} we present the 
	non-linear Compton matrix elements in strong-field QED in the Furry picture.
	Special emphasis is put on a precise definition of the spin-polarization properties of asymptotic states as well as their time-evolution
	in external fields as described by the Volkov states.
	Section \ref{sect:densitymatrix} is devoted to the definition of the polarization density matrix of the Compton scattered electrons as central object.
	Numerical investigations for a few examples of the polarization properties of Compton scattered electrons are presented in 
	Section \ref{sect:numerics}.
	In Section \ref{sect:analytic} we derive the LCFA approximation of the polarization density matrix.
	We find explicit expressions for spin-flip rates and non-flip rates for arbitrary initial electron polarization.
	The LCFA approximation is compared quantitatively with exact QED results.
	Our results are discussed and summarized in Sect.~\ref{sect:summary}.
	Technical details have been relegated to two Appendices.

	\section{Theoretical Background}

	\label{sect:theory}

	\subsection{Volkov States and the Strong-Field S-Matrix}

	The theory of strong-field QED in high-intensity laser fields,
	including the nonlinear Compton scattering process \cite{Nikishov:JETP1964a,Nikishov:JETP1964b,
	Goldman:PhysLett1964,Brown:PR1964,Nikishov:JETP1965,Narozhnyi:JETP1965,Seipt:PRA2011}
	is based on the Furry picture, where the interaction of the electrons with a background laser field $A$ is
	treated non-perturbatively by using laser dressed 	Volkov states \cite{Volkov:ZPhys1935}.
	This treatment is possible when the laser pulse is described as a uni-variate plane-wave external field 
	$A^\mu = A^\mu(\phi)$ which is a sole function of the laser phase $\phi = k.x$,
	with light-front gauge $k.A = 0$ and the light-like four-wavevector $k^2=0$,
	for generalizations see e.g.~Refs.~\cite{Raicher:PRA2013,King:PRD2016,Heinzl:PRD2016}.
	The Volkov states $\Psi(x)$ are solutions of the minimally coupled Dirac equation
	\begin{align}
 	 \left( i\slashed\partial - \slashed A(\phi) + m \right) \Psi(x) = 0 \,,
  	\label{eq:dirac}
	\end{align}
	where $m$ is the electron mass and we absorbed the charge $e$ into the vector potential $eA\to A$.
	We employ the Feynman slash notation $\slashed A=\gamma.A  = (\gamma A)= \gamma^\mu A_\mu$ for
	scalar products of four-vectors with the Dirac matrices $\gamma^\mu$.
	Moreover, throughout the paper we use Heaviside-Lorentz units with $\hbar=c=1$.

	An explicit representation of the Volkov wave function \cite{Volkov:ZPhys1935} is given by
	\begin{align} \label{eq:Volkov}
	\Psi_{ps}(x) &= \left( 1 + \frac{\slashed A \slashed k}{2 k.p } \right) \, e^{-i S_p(x)} \,  u_{ps}\,, \\
	S_p(x) &= p.x + \int \! \frac{\d \phi}{k.p} \left[ p.A - \frac{A.A}{2} \right] \,,
	\end{align}
	for an electron with asymptotic four-momentum $p$, i.e. before the electron interacts with the laser pulse.
	Here, $u_{ps}$ denotes the usual free Dirac spinor,
	and $S_p(x)$ is the classical Hamilton-Jacobi action of the electron in the background field $A$.
	Moreover, $s = \pm 1$ denotes the spin projection quantum number of the electron,
	which will be discussed in more detail below in Section \ref{sect:spin}.

	By choosing the initial conditions for the solution of the Dirac equation in the background field we can make sure that the
	Volkov states \eqref{eq:Volkov} turn into free Dirac wave functions at asymptotic past for incoming particles,
	$\Psi_{ps}(x) \stackrel{\phi\to -\infty}{\longrightarrow} e^{-ip.x} u_{ps}$,
	when the electron is separated from the field.
	Similarly we argue for the adjoint Volkov wave functions for outgoing electrons that
	$\overline{\Psi}_{p's'}(x) \stackrel{\phi\to+\infty}{\longrightarrow} \bar u_{p's'}e^{ip'.x}$.
	These choices ensure that the momentum $p$ and spin quantum number $s$ actually describe the electron's properties
	outside the strong laser pulse which can be observed by a detector \cite{Ilderton:PRD2013}.

	The S-matrix element describing the emission of a
	photon with four-momentum $k'$ in the non-linear Compton process is given by \cite{book:Landau4,Mitter:ActaPhysAustr1975,Ritus:JSLR1985,Harvey:PRA2009,
		Seipt:PRA2011,King:PRA2013,Seipt:JPP2016}
	\begin{align}
	S & 	=
		  -i e \intop \! \ud^{4}x \: \overline{\Psi}_{p's'}(x) \,  
		  		\slashed \varepsilon'^* e^{i k'.x} \, \Psi_{ps}(x) \,,
	\label{eq:Smatrix}
	\end{align}
	where $\varepsilon'^*_\mu e^{i k'.x}$ is the wave function of the emitted photon.

	The space-time integrations in \eqref{eq:Smatrix} are easiest performed using
	light-front coordinates, $x^\pm = x^0 \pm x^3$ and $\vec x^\perp = (x^1,x^2)$ and,
	after performing the spatial integrals over $\ud x^-$ and $\ud^2\vec x^\perp$ the S matrix can be written written as
	\begin{align}
	S & =
	- i e (2\pi)^{3} \:
	\delta_\mathrm{l.f.}( \mathsf{k}' + \mathsf{p}' - \mathsf{p} ) \:
	\bar u_{p's' } \, \mathcal M \, u_{p s } \,,
	\label{eq.S_matrix.final}
	\end{align}
	where $\mathsf p \equiv (p^+ ,\vec p^\perp)$, and the light-front delta function
	$\delta_\mathrm{l.f.}( \mathsf{k}' + \mathsf{p}' - \mathsf{p} ) 
	= \frac{2}{k^-}\delta^{2}(\vec{k}'^\perp+\vec{p}'^\perp - \vec{p}^\perp) \delta(k'^+ + p'^+ - p^+ )$
	ensures the conservation light-front momentum during the scattering.

	The amplitude $\mathcal M$ contains an integral over the laser phase $\phi = \omega x^+$,
	\begin{align}
	\mathcal M
	= \varepsilon'^*_\mu \int \! \ud \phi \: \mathscr J^\mu(\phi)  \: e^{i \int^\phi \! \ud \phi' \: \frac{k' . \pi(\phi') }{ k.p'}} \,,
	\label{eq:matrix.element} 
	\end{align}
	with
	\begin{align} \label{eq:current-integrand}
		\mathscr J^\mu(\phi) &= \gamma^\mu 
			+ \frac{\slashed A \slashed k \gamma^\mu }{2k.p'} +  \frac{ \gamma^\mu \slashed k \slashed A}{2k.p}
			 + \frac{\slashed A \slashed k \gamma^\mu \slashed k \slashed A}{4 (k.p)(k.p')} \,.
	\end{align}
	and with the classical kinematic four-momentum of the electron
	\begin{align} \label{eq:pi}
	\pi^\mu(\phi) &= p^\mu - A^\mu(\phi) + k^\mu \, \frac{A.p}{k.p} - k^\mu \, \frac{A.A}{2k.p} \,.
	\end{align}
	The latter is a solution of the classical Lorentz force equation
	$m \frac{\mathrm d \pi^\mu}{d\tau} = F^{\mu\nu} \pi_\nu$ ($\tau$ denotes proper time here)
	in the given background field with Faraday tensor $F^{\mu\nu} = \partial^\mu A^\nu - \partial^\nu A^\mu$.

	Here it becomes clear again that $p^\mu$ refers to the asymptotic value of the momentum when $A\to 0$.
	We mention that neither Eq.~\eqref{eq:current-integrand} nor Eq.~\eqref{eq:pi} depend on
	the polarization of the involved electrons and the emitted Compton photon.
	All polarization dependence is contained in the electron Dirac spinors $u$ and $\bar u$, as well as
	the photon polarization vector $\varepsilon'$.

	\subsection{Description of the Spin of Volkov Electrons}
	\label{sect:spin}

	Here we describe in detail the how the Dirac spinors $u_{ps}$ encode the spin properties of the electrons,
	we will give a precise meaning to the spin quantum number $s$. The Volkov states describe electrons that
	are subjected to a background field. We therefore split the discussion into two parts, discussing free electrons
	first and afterwards generalizing to electrons in a background field.

	Because we are interested in solutions of the Dirac equation with well-defined spin polarization states
	we need to construct them as eigenstates of a spin operator that is also a constant of motion \cite{book:Sokolov}.
    However, the components of the relativistic (Pauli) spin operator
	$\vec \Sigma =  \gamma^5 \gamma^0 \vec\gamma $ do not commute with the free relativistic
	Dirac Hamiltonian $H = \gamma^0 \vec \gamma \cdot \vec p  +  \gamma^0 m $.
	In the non-relativistic limit, the corresponding spin-operators---the Pauli-matrices $\vec \sigma$---do commute with the
	free non-relativistic Hamiltonian. It is therefore customary to describe the spin of a relativistic particle in its rest frame,
	where we can apply the non-relativistic theory \cite{book:Landau4}, see also Refs.~\cite{Fradkin:RevModPhys1961,Bauke:NJP2014,Bliokh:PRA2018}
	a general discussion and for alternative approaches.

	The components of neither the non-relativisitic or relativisitic spin operators mutually commute,
	$[\sigma_i,\sigma_j] = 2i\epsilon_{ijk} \sigma_k$.
	That means only one component of spin can be measured at any time, i.e.
	we need to choose a quantization axis $\vec \zeta$ ($|\vec\zeta| = 1$) for the spin. 
	Following \cite{book:Landau4},
	we define the free Dirac bi-spinor in the rest frame of the electron as
	\begin{align} \label{eq:spinor-rf}
	u_{\vec \zeta s}^\mathrm{RF} = 
	\left(
	\begin{matrix}
 	w_{\vec\zeta s} \\
 	0 
	\end{matrix}
	\right) \,,
	\end{align}
	with a (non-relativistic) two-component spinor $ w_{\vec\zeta s}$
	\footnote{Explicit expressions for $ w_{\vec\zeta s}$ can be obtained from the standard
	representation $w_{\vec e_z s} = (\delta_{s,1},\delta_{s,-1})$
	by rotating them via the Wigner $D$ functions \cite{book:Balashov}.
	For clarity we label the spinors from now on according to their spin quantization direction.}.
	The latter is an Eigenfunction of the spin-projection along the quantization direction $\vec \zeta$ , i.e.~of the operator $(\vec\zeta \cdot \vec \sigma)$,
	\begin{align}
	(\vec\zeta \cdot \vec \sigma)  w_{\vec\zeta s}  = s w_{\vec\zeta s}\,,
	\end{align}
	and which is normalized according to $w_{\vec\zeta s}^\dagger w_{\vec\zeta s'} = \delta_{ss'}$.
	That means the two-component spinors $w_{\vec\zeta s}$, and therefore also $u_{\vec\zeta s}^\mathrm{RF}$,
	describe electrons with their spin polarized
	along the axis $\vec \zeta$ in the rest frame, with the quantum number $s=+1$ $(-1)$ denoting the spin
	is aligned parallel (antiparallel) to the axis $\vec \zeta$.
	The matrix elements of the Pauli-matrices between the states $w_{\vec\zeta s}$ are given by
	\begin{align}
	 w^\dagger_{\vec\zeta s'} \vec \sigma w_{\vec\zeta s}  = 
	\left( 
	\begin{matrix}
	\vec \zeta & \vec e_- \\
	\vec e_+ & -\vec \zeta
	\end{matrix}
	 \right)_{s's}  \equiv \vec S_{ss'}\,, \label{eq:wsw}
	 \end{align}
	with $\vec S_{ss} = s\vec\zeta$, $\vec S_{\up\down} = \vec e_{+}$ and $\vec S_{\down\up} = \vec e_{-} $.
	The vectors on the off-diagonal elements are perpendicular to $\vec\zeta$, and defined as
	$\vec e_\pm = \vec e_1 \pm \vec e_2$, where
	$\vec e_1 = (\cos \vartheta \cos \varphi , \cos \vartheta  \sin \varphi , - \sin \vartheta)$
	and $\vec e_2 = (-\sin\varphi,\cos\varphi,0)$ with 
	$\vartheta$ and $\varphi$ as the polar and azimuthal
	angles of the spin quantization direction, respectively, i.e.~$\vec \zeta = ( \sin\vartheta \cos \varphi , \sin\vartheta \sin\varphi , \cos \vartheta)$.
	Similarly, we find for the dyadic products of the spinors
	$w_{\vec \zeta s} w^\dagger_{\vec \zeta s'} = ( \delta_{ss'} + \vec \sigma \cdot \vec S_{ss'}) / 2 $.
	Its relativistic generalization appears in the calculations of the squared S matrix elements.

	The relativistic Dirac bi-spinors in the laboratory frame $u_{p\vec\zeta s}$ (which are appear in the Volkov states)
	are solutions of the algebraic equation $(\slashed p - m ) u_{p\vec \zeta s} =0 $. 
	This condition is always fulfilled if we construct them in the following way \cite{book:Itzykson}
	\begin{align} \label{eq:up}
	u_{p\vec\zeta s} = \frac{\slashed p + m}{\sqrt{E+m}} \: u_{\vec \zeta s}^\mathrm{RF} 
	\end{align}
	for the on-shell four-momentum  $p^\mu \equiv (E,\vec p)$, and with the rest frame spinors \eqref{eq:spinor-rf}.
	Those spinors are normalized as $\bar u_{p\vec \zeta s} u_{p\vec \zeta s'} = 2m \delta_{ss'}$.
	In addition, we can easily verify the following vector and axial-vector bi-linear forms of the bi-spinors
	\begin{align}
	\bar u_{p\vec\zeta s} \gamma^\mu u_{p\vec\zeta s} &= 2 p^\mu 
		\label{eq:bilinear:vector}\,,\\
	 \bar u_{p\vec\zeta s} \gamma^\mu \gamma^5  u_{p\vec\zeta s} 
		&=  2m s \zeta^\mu  \,,
		\label{eq:bilinear:axialvector}
	\end{align}
	Here, we introduced the covariant spin four-vector
	\begin{align}
	\zeta^\mu & \equiv  \left( \frac{\vec\zeta \cdot \vec p}{m} \, , \:  \vec\zeta + \frac{\vec p (\vec\zeta\cdot \vec p)}{m(E+m)} \right)\,,
	\end{align} 
	which is the Lorentz transform of the spin quantization axis $\vec \zeta$ defined in the rest frame of the electron,
	$\zeta^\mu = \Lambda^\mu_{\ \nu} ( \vec p) \zeta^\nu_\mathrm{RF}$, where
	$\zeta^\nu_\mathrm{RF} \equiv (0,\vec\zeta)$ \cite{book:Jackson,book:Landau4}.
	We note that $\zeta^\mu$ is a space-like axial unit four-vector, $\zeta.\zeta = -1$,
	that is perpendicular to the particle's momentum, $\zeta.p=0$.
	We can also show that the covariant spin four-vector is related to the expectation value of the
	Pauli-Lubanski pseudo-vector $W^\mu = -\frac{1}{4} \epsilon^{\mu\alpha\beta\tau} \sigma_{\alpha\beta} p_\tau$,
	namely $\bar u_{p \vec \zeta s} W^\mu u_{p \vec\zeta s} = m^2 \, s \zeta^\mu$.
	Here $\sigma_{\alpha\beta} = \frac{i}{2} [\gamma_\alpha,\gamma_\beta]$ is the
	commutator of the Dirac matrices and $\epsilon^{\mu\nu\alpha\beta}$ the completely antisymmetric Levi-Civita tensor.

	Later we will also need the outer products of the Dirac bi-spinors of the form \cite{Lorce:PRD2018}
	\begin{align} \label{eq:uubar}
	u_{p\vec \zeta s}         \bar u_{p \vec \zeta s'    }  		  
		= \frac{1}{2} (\slashed p + m ) (\delta_{ss'} -  \gamma^5 \slashed S_{ss'} ) \,,
	\end{align}
	with $S_{ss} = s \zeta $, $S_{\uparrow\downarrow} = e_+$, and $S_{\downarrow\uparrow}=e_-$,
	and where $e_\pm$ are the Lorentz transforms of the vectors $\vec e_\pm$ defined above in the rest frame of the electron. 
	We note that the matrix $S_{ss'}$ is called the covariant spin-density matrix in Lorc\'e \cite{Lorce:PRD2018}.

	Now that we established all necessary relations for the free Dirac bi-spinors let us turn our attention to the problem of the 
 	interpretation of the  above spin polarization properties when the electron interacts
	with a strong electromagnetic background field.
	To this end, let us first calculate the vector and axial vector expectation values between Volkov states.
	By using the bi-spinors \eqref{eq:up} in \eqref{eq:Volkov},
	and together with \eqref{eq:bilinear:vector} and \eqref{eq:bilinear:axialvector} it is easy to 
	show that the bi-linears of Volkov states are \cite{Ritus:JSLR1985}
	\begin{align}
	\bar \Psi_{p\vec\zeta s} 											\Psi_{p\vec\zeta s} & = 2 m \,, \\
	\bar \Psi_{p\vec\zeta s} \gamma^\mu 						\Psi_{p\vec\zeta s} & = 2  \pi^\mu(\phi) \,, \\
	\bar \Psi_{p\vec\zeta s}  \gamma^\mu \gamma^5	\Psi_{p\vec\zeta s} & =  2 m \, \Sigma^\mu(\phi) \,.
	\end{align}

	Here, $\pi^\mu(\phi)$ denotes again the classical kinematic four momentum, Eq.~\eqref{eq:pi},
	which is the solution of the classical Lorentz force equation. 
	The asymptotic momentum $p^\mu$---which labels the Volkov states---serves hereby as the initial value
	of the solution of the orbital equation of motion. We note also that all laser pulses considered are characterized by $A^\mu(\phi \to \pm \infty) \to 0$. 	%
	Moreover, $\Sigma^\mu$ is the solution of the Bargman-Michel-Telegdi equation (with gyromagnetic ratio $g_e=2$)
	\cite{Bargmann:PRL1959,Spohn:AnnPhys2000}, $\frac{\ud \Sigma^\mu}{\ud \tau} = \frac{g_e}{2} F^{\mu\nu} \Sigma_\nu$,
	that describes the classical motion of a magnetic moment in a given background field, and with the initial conditions given by $ s\zeta^\mu$. 
	That means, by working with Volkov states the classical spin-precession of the
	electrons due to the interaction with the laser field is already included in the calculations.
	The quantum numbers $s$ and the direction $\vec \zeta$ ($s',\vec\zeta'$) describe the asymptotic properties of the incident and
	final particles when they are separated from the field and are indeed observables.

	The explicit solution of the BMT equation with initial condition $\Sigma^\mu(-\infty) = S^\mu$ is given by
	\begin{align} \label{eq:BMT-solution}
	\Sigma^\mu(\phi) &= S^\mu - A^\mu \frac{(k.S)}{k.p} 
										+ k^\mu \frac{(A.S)}{k.p}
										- k^\mu \frac{(A.A) (k.S) }{2(k.p)^2} \,.
	\end{align}
	It is easy to shows that for each $\phi$, Eq.~\eqref{eq:BMT-solution} is orthogonal to the kinematic four-momentum $\pi^\mu$, $\pi.\Sigma = 0$.
	Moreover, $\Sigma^\mu(+\infty) = \Sigma^\mu(-\infty)$ for any admitted plane wave laser field $A^\mu$.

	\subsection{A Physical Spin Basis}
	\label{sect:basis}
	
	For analytical calculations involving electrons in a strong field it is often convenient to choose a special basis vector for the spin quantization,
	\begin{align}
	\label{eq:spin-basis-lab}
	\zeta^\mu &= \beta^\mu - \frac{p.\beta}{k.p} \: k^\mu \,,
	\end{align}
	for the initial state electron and accordingly $p$ replaced by $p'$  for the vector $\zeta'$ of the final state electron \cite{King:PRA2015}.
	While $k^\mu \equiv  ( \omega , \omega \hat{\vec k})$, with frequency $\omega$ and propagation direction $\hat{\vec k}$, is the (light-like) four-wavevector of the
	strong background field,
	$\beta^\mu \equiv (0,\vec \beta)$, with $|\vec \beta|=1$, denotes the direction of the magnetic field in the laboratory frame.

	As we have seen above, the choice of $\zeta$ as the spin quantization axis
	induces the canonical spin basis $\{ \zeta,e_1,e_2\}$ in the dyadic products of the Dirac spinors.
	However, this canonical basis is not very convenient for explicit analytical calculations.	
		
	Instead, we now define a physically motivated basis of three space-like four-vectors $\{\zeta,\eta,\kappa\}$
	in which analytical calculations become particularly simple, and we show its relation to the canonical basis.
	In addition to $\zeta$ in \eqref{eq:spin-basis-lab} we define (here for the incident electron)
	\begin{align}
	\label{eq:spin-basis-lab-add}
	\begin{split}
	\eta^\mu   &= \varepsilon^\mu - \frac{p.\varepsilon}{k.p} \: k^\mu \,, \\
	\kappa^\mu &= \frac{m k^\mu}{k.p} - \frac{p^\mu}{m} \,,
	\end{split}
	\end{align}
	where $\varepsilon^\mu \equiv  (0,\vec \varepsilon)$ denotes the direction of the electric field in the laboratory frame.
	The three vectors $\{\zeta,\eta,\kappa\} $ are mutually orthogonal, space-like unit four-vectors
	(e.g.~$\zeta.\zeta = \eta.\eta =-1$, $\zeta.\eta=0$ etc.)
	which are also orthogonal to the electron momentum $p$, 
	and by means of 
	\begin{align} \label{eq:eps-times-beta}
	\hat{ \vec k }  = \vec \varepsilon \times \vec \beta
	\end{align}
	they form a right-handed basis.

	These three vectors, Lorentz boosted to the (asymptotic) rest frame of the particle,
	{$\{ \zeta,\eta,\kappa\} \stackrel{\Lambda(-\vec p)}{\longrightarrow} 
	\{ \vec \zeta_\mathrm{RF}, \vec \eta_\mathrm{RF},  { \vec \kappa}_\mathrm{RF}\}$}, become
	exactly the directions of the magnetic field, electric field, and wave vector of the background field in the electron's rest frame,
	e.g.
	\begin{align}
	\vec \zeta_\mathrm{RF} = \frac{\vec B_\mathrm{RF}}{| \vec B_\mathrm{RF}|}
	=
	\vec \beta + \frac{ \vec\beta\cdot \vec p  }{ p^+ } \left( \hat{\vec k} - \frac{\vec p}{m+E} \right)
	\end{align}
	according to the transformation laws of the electromagnetic field, cf.~e.g.~\cite{book:Jackson},
	and by using $|\vec B| = |\vec E|$ for a plane wave field, as well as Eq.~\eqref{eq:eps-times-beta}.
	The expression for $\vec \eta_\mathrm{RF}$ is similar but with $\vec \beta\to\vec\varepsilon$, and 
	$\vec \kappa_\mathrm{RF} = \vec \eta_\mathrm{RF} \times \vec \zeta_\mathrm{RF} $.

	We now have two different sets of vectors perpendicular to $\zeta$.
	Those two bases are related by a rotation in the plane perpendicular to $\zeta$, written in complex form as 
 	\begin{align} \label{eq:phaseangle+}
	e_+ 	= e^{i\Phi} ( \kappa +i\eta )  \,.
 	\end{align}
 	We shall show now that, under typical conditions for an electron laser pulse collision the rotation angle $\Phi$ is small.
	
	In the typical scattering scenario we envisage here, the electron initially counter-propagates relative to the intense laser pulse with $p^+ \gg m$
	(i.e. Lorentz factor $\gamma \gg 1$), $\xi \gg 1$ and $m\xi \ll p^+$, and $p_\perp \ll p^+ $.
	We can express the rotation angle (for the incident electron) as
	$\sin \Phi =  \vec \kappa_\mathrm{RF} \cdot  \vec e_2 
	= \vec \kappa_\mathrm{RF}\cdot  \vec \varepsilon \sin\varphi +  \vec \kappa_\mathrm{RF} \cdot   \vec \beta \cos\varphi$,
	where $\varphi$ is the azimuthal angle of $\vec \zeta_\mathrm{RF}$, which
	is given by
	$$\cot \varphi =   \frac{ \vec \zeta_\mathrm{RF} \cdot \vec \varepsilon}{\vec \zeta_\mathrm{RF}\cdot \vec\beta}
	\simeq \frac{ (\vec p\cdot \vec\beta) (\vec p \cdot \vec \varepsilon)}{ p^+  (m+E)} 
	= \mathcal O \left( \frac{p_\perp^2 }{ p^+ E }\right) \ll 1 \,.$$
	With this we can express 
	$\sin\Phi = \mp  \frac{(\vec{\hat  k} \cdot \vec \eta_\mathrm{RF} )}{\sqrt{1 - ( \vec{\hat k} \cdot \vec \zeta_\mathrm{RF})^2}}$,
	where $\vec{\hat  k} \cdot \vec \eta_\mathrm{RF},\vec{\hat  k} \cdot \vec \zeta_\mathrm{RF} = \mathcal O( p_\perp / p^+\gamma)\ll1$
	and therefore $ \Phi \ll 1$.

	Repeating the calculation for the final electron we find
	$\sin \Phi' \simeq \frac{(\vec p_\perp - \vec k'_\perp)\cdot \vec \varepsilon}{(1-t)p^+ E'} 
	= \mathcal O (\frac{m^2\xi}{(1-t)^2(p^+)^2})\ll 1$
	for $1- t \gg  \sqrt{\xi} m /  p^+  $.
	While for $\chi_e\ll 1$ the typical values of $t \sim \chi_e$ are small, for $\chi_e \gg1$ 
	the probability distribution has a maximum at $t = 1 - 4/3\chi_e$ with an exponential fall-off beyond that point \cite{Bulanov:PRA2013}.
	Thus, we can safely approximate to leading order  $e^{i\Phi'} \simeq 1 $
    in the former case always,
	and in the latter case if $m\sqrt{\xi} \chi_e /p^+  \ll 1$ is fulfilled, i.e. $\xi \ll (m/\omega)^{2/3}$. 
	For electrons interacting with a typical Ti:Sa laser with $\omega=\unit{1.55}{\electronvolt}$ the
    approximation is valid as long as $\xi \ll 4700$.

	\subsection{Polarized and Unpolarized Scattering Probability}

	Having clearly defined the meaning of the spin polarization indices of the spinors in Sect.~\ref{sect:spin}
	we can write the scattering probability by squaring the S matrix elements \eqref{eq:Smatrix}.
	To be very clear, we will here explicitly write the (completely arbitrary)
	quantization directions $\vec \zeta$ and $\vec \zeta'$ for the initial and final electrons,
	and obtain for the polarized probability
	\begin{align} \label{eq:probability-pol}
	\mathbb P(s,\vec \zeta ; s' , \vec \zeta' ; \lambda' ) 
	= \frac{1}{2p^+} \int | S (s,\vec \zeta ; s' , \vec \zeta' ; \lambda')|^2 \, 	\widetilde{\d\vec k}' \widetilde{\d\vec p}' \,
	\end{align}
	where $\widetilde{\d\vec p}' = \frac{\d^2 \vec p'_\perp \d p'^+ }{(2\pi)^3 2 p'^+} $
	and similarly $\widetilde{\d\vec k}'$ denote the Lorentz invariant phase space elements of the final state
	electron and photon, respectively. The leading factor $1/2p^+$ is the flux factor of the initial electrons.
	The expression \eqref{eq:probability-pol} corresponds to the probability that a photon with helicity $\lambda'$ is emitted
	by an electron that was polarized with projection $s = \pm1$ along the direction $\vec \zeta$ before the scattering,
    while the electron's
	polarization is measured simultaneously with projection $s' = \pm1$ along the direction $\vec \zeta'$ after the scattering.

	By using \eqref{eq:probability-pol} it is straightforward to determine also the probability for
	unpolarized particles, i.e. when the polarization of none of the particles is known or measured,
	by summing over all final state polarizations and averaging over initial state polarization of
	the electron:
	\begin{align} \label{eq:probability}
	\mathbb P = \frac{1}{2} \sum_{\lambda',s',s} \mathbb P (s,\vec\zeta; s' ,\vec \zeta' ,\lambda') \,.
	\end{align}
	It is clear that this probability must be independent of the choice of the bases $\vec \zeta$ and $\vec \zeta'$ in which
	we measure the polarization properties of the electrons. 
	In order to calculate the scattering probabilities also for \textit{partially polarized} electrons, and for determining
	the state of polarization of the scattered electrons it is convenient to use the density matrix formalism which we
	introduce now.

	\section{The Polarization Density Matrix of Scattered Electrons}

	\label{sect:densitymatrix}

	So far we only discussed spin on the level of wave functions and amplitudes. But within this framework it is possible to describe only
	pure states, i.e. completely polarized states. In order to describe mixed (or partially polarized) states we now  introduce the
	spin density matrix \cite{book:Blum,book:Baier}.
	This density matrix framework is most convenient to describe polarization phenomena in many areas of physics,
	see e.g.~Refs.~\cite{book:Balashov,Surzhykov:RadPhysChem2005,Lotstedt:PRL2009,Kampfer:PRA2016,Peshkov:PRA2017},
	and we use it in the following to describe the radiative polarization of the electrons during non-linear Compton scattering.

	The density operator of the final state of the scattered particles $\hat \rho_f$
	is related to the density operator of the initial state $\hat \rho_i$ by the relation
	\begin{align}
	\hat \rho_f = \hat S \: \hat \rho_i \: \hat S^\dagger\,,
	\end{align}
	where $\hat S$ denotes the S operator (in the Furry picture) for the considered scattering process.
	In our work we use Eq.~\eqref{eq:Smatrix}.
	The density operators $\hat \rho_f$ and $\hat \rho_i$ contain the full information of the final and initial states, respectively.

	Usually one is not interested in the full density matrix in which case one has to calculate
	\textit{reduced} density matrices by taking the trace over unobserved properties \cite{book:Blum}.
	For (discrete) polarization indices this means summing over all polarization states analogously to
	\eqref{eq:probability}, while for continuous momentum variables this means integration over final state phase space.

	For the initial state density operator $\hat \rho_i$ (describing the state of the initial electron) we may assume that
	it can be represented as a direct product of a momentum state density matrix and a polarization density matrix
	$\hat \rho_i = \hat \rho_{i}^\mathrm{(momentum)} \otimes \hat \rho_{i}^\mathrm{(pol)}$ if these degrees of freedom
	are uncorrelated before the interaction.
	The polarization density matrix of the initial electrons is just a $(2\times2)$ matrix \cite{book:Balashov}, which can be represented as
	\begin{align} \label{eq:rho-spin}
	\rho_i^{(\mathrm{pol})} = \frac{1}{2} (1 +  \vec \sigma \cdot \vec \Xi  )\,,
	\end{align}
	with the Pauli matrices $\vec \sigma$ and the \Stokes vector $\vec \Xi$ of the initial electrons,
	and with $\tr \rho_i^{(\mathrm{pol})} = 1$.

	The \Stokes vector $\vec \Xi$ describes both the direction and the degree of polarization of the electrons $ | \vec \Xi |$.
	In case of $  | \vec \Xi | = 1$ the density matrix \eqref{eq:rho-spin} describes a fully polarized pure state.
	For $0  \leq  | \vec \Xi | < 1$ the initial electrons are in a partially polarized mixed state, and,
	in particular for $ | \vec \Xi | = 0$ the initial electrons are unpolarized.

	For the momentum state
	we assume that the electrons are in a pure plane wave state before the interaction,
	such that we can just take $ \rho_{i}^\mathrm{(momentum)} = 1/ 2p^+$ which is the flux factor of the initial electron.
	Note that the initial electron density matrix is conveniently normalized to said flux factor $\tr \rho_{i} = 1/2p^+$.
	It would be straightforward to include here also the scattering of electron wave packets \cite{Stock:PRA2015,Angioi:PRA2016}.

	The reduced polarization density matrix of the final electrons is given by
	the following expression (we suppress the subscript $f$ for the final density matrix)
	\begin{widetext}
	\begin{align}  \label{eq:rho-final-reduced}
	\rho_{\bar s' s'}
		 &= \int \widetilde{\d\vec k}' \widetilde{\d\vec p}' \: 
		 			\sum_{s \bar s , \lambda'} 
		 				S(s,\bar s',  \lambda') 
		 				\, (\rho_{i})_{s \bar s} \,
		 				S^\dagger(\bar s , s' ,\lambda ' ) 
			 = 		 \intop_{\mathbb R^2} \! \d^2\vec k'_\perp 
			 			\int_0^1 \! \ud t \: \rho_{\bar s' s'} ( \vec k'_\perp, t) 
	\end{align}
	with
	\begin{align}
	\rho_{\bar s' s'}( \vec k'_\perp, t)  &= 	- \frac{ \alpha  }{16\pi^2 (k.p)^2 t(1-t) } 		
			\sum_{s\bar s} \left( \rho_{i }^\mathrm{(pol)} \right)_{s\bar s}
			\intop \! \d\phi \d\phi' 
			 \:  e^{i\int_{\phi'}^\phi \! \ud \tilde \phi \,  \frac{k'. \pi (\tilde \phi )}{k.p'}}
			\: 
			 (\bar u_{p' \bar  s'} \mathscr J^\mu(\phi) u_{ps}) \: ( \bar u_{p \bar s} \bar{\mathscr J}_\mu(\phi')  u_{p's'}) 
			 \label{eq:rho-differential}
	\end{align}
	\end{widetext}
	with Sommerfeld's fine structure constant $\alpha = e^2 /4\pi$ and $t = (k.k')/(k.p)$.
	Because we are not interested in the polarization of the emitted photons, we summed over the
	final photon polarization indices $\lambda'$, and applied the well known relation for the photon polarization vectors 
	$\sum_{\lambda'}\varepsilon_{\mu}(\lambda') \varepsilon^*_{\nu}(\lambda') = - g_{\mu\nu}$ \cite{book:Itzykson}.

	The expression \eqref{eq:rho-final-reduced} above refers to a reduced density matrix where
	we integrated over all momenta of the final state particles, and which is normalized to the total scattering probability,
	$\tr \rho = \mathbb P$, in full agreement with~\eqref{eq:probability} (we explicitly show this below in Section
	\ref{sect:analytic}).
	The integrand of \eqref{eq:rho-final-reduced}, $\rho_{\bar s' s' }( \vec k'_\perp, t)$, given in
	\eqref{eq:rho-differential} refers to a momentum differential reduced density matrix, 
	which is normalized accordingly to the differential scattering probability $\tr \rho (\vec k'_\perp, t) = \frac{\ud \mathbb P}{\ud t \ud^2 \vec k_\perp'}$.
	In general, whenever we explicitly note the momentum dependence of a density matrix, it is always 
	normalized to the corresponding differential probabilities.

	The polarization of the scattered electrons is obtained by calculating the final electron \Stokes vector by using
	\eqref{eq:rho-final-reduced}, e.g.~via
	\begin{align}
	\vec \Xi' = \frac{ \tr [ \vec\sigma \rho ] }{\tr \rho} 
	\,,\qquad 
	\vec \Xi'(\vec k'_\perp, t) = \frac{ \tr  [ \vec\sigma \rho(\vec k'_\perp, t) ] }{\tr \rho(\vec k'_\perp, t)} 
	\end{align}
	for the total and differential version.
	By having chosen an arbitrary quantization axis $\vec \zeta'$ for the outgoing spinors,
	the density matrix constructed in this section is diagonal in $\vec \zeta'$,
	i.e. the projection with $\sigma_3$ determines the polarization along $\vec \zeta'$.
	Similarly, the projections with $\sigma_{1,2}$ give the polarizations along $\vec e'_{1,2}$.

	\section{Physical Example Calculations}
	\label{sect:numerics}

	As we argued above, the choice of the spin-basis is irrelevant for the calculation of the scattering probability and
	the final electron \Stokes vector.
	For the numerical investigations in this section we conveniently 
	choose $\vec \zeta' = \vec \zeta = \vec e_z= (0,0,1)$ for both the initial and final electrons,
	for which $\vec e_1 = \vec e_x$ and $\vec e_2 = \vec e_y$.
	For the numerical integration of the highly oscillating integrals over the laser phase in \eqref{eq:rho-differential} we
	adapted the algorithm outlined in Ref.~\cite{Thomas:PRSTAB2010}.

	\subsection{Even On-Axis Harmonics, Spin-Flips and Angular Momentum Conservation}

	Let us begin our numerical analysis by investigating with an example where can nicely observe the transfer of
	spin angular momentum from the laser beam to the electrons (cf.~also \cite{Ahrens:PRA2017}).
	We consider the head-on scattering of an electron
	beam with $\gamma=5000$ and $\vec p_\perp=0$ on a moderately intense linearly polarized laser pulse with $\xi = 0.5$
	and central frequency $\omega = \unit{1.55}{\electronvolt}$.
	That means the quantum energy parameter $b =  0.03$ and $\chi_e = 1.5\times10^{-2}$.
	The vector potential is $A^\mu = m\xi \varepsilon^\mu h(\phi)$,
	with the dimensionless shape function of the laser vector potential having a Gaussian envelope
	$h(\phi) = e^{-\phi^2/2\Delta\phi^2}\cos \phi $ with rms width $\Delta\phi = 50$.

\begin{figure}[tb]
	\includegraphics[width=0.99\columnwidth]{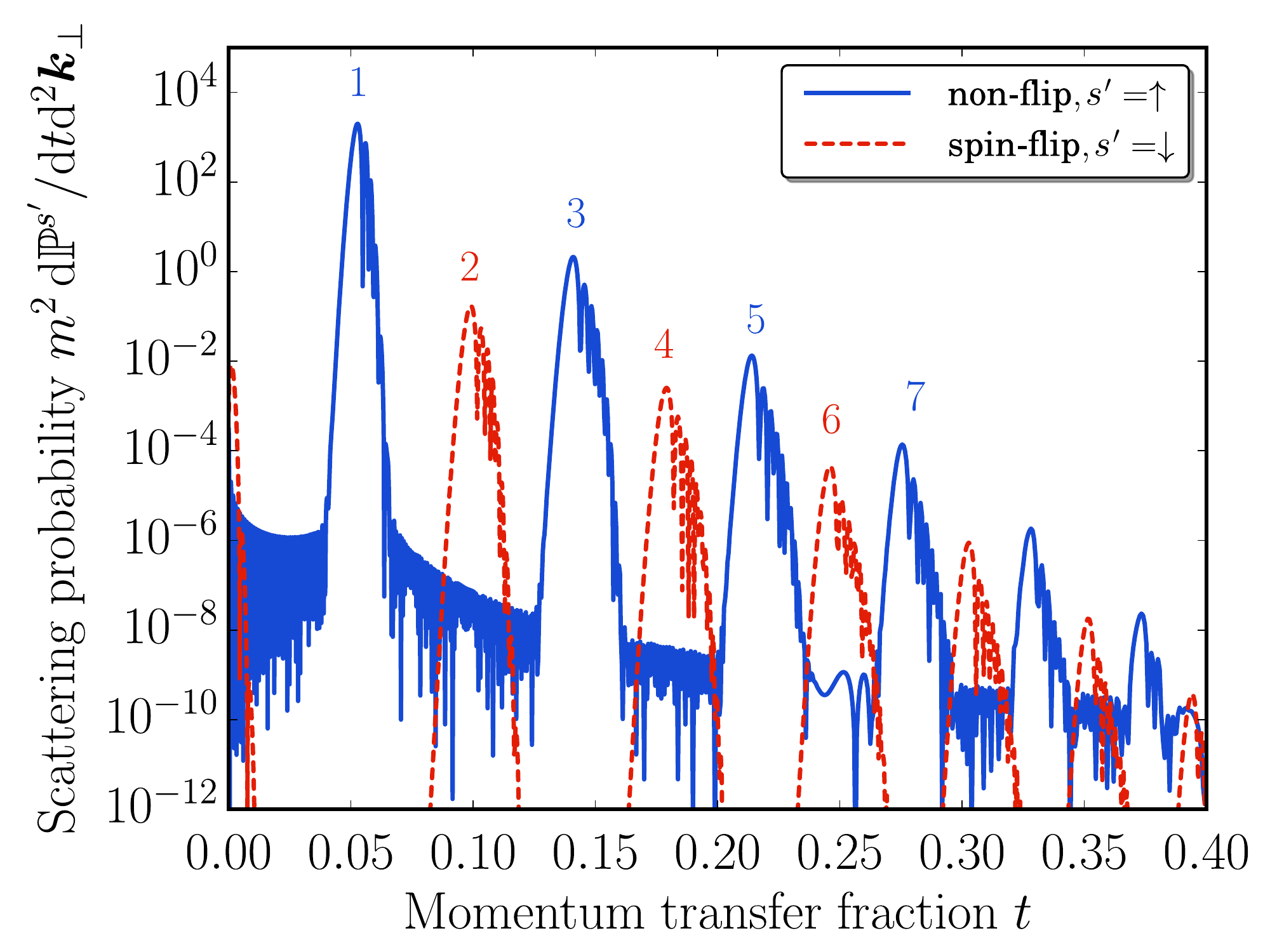}
	%_1d\_spectra/input20
	\caption{On-axis spectrum for the head-on collision of a polarized electron beam, initially polarized along the beam axis.
	The even harmonics are weaker than the odd harmonics, and they are due to spin-flip transitions.
	For parameters see text.}
	\label{fig:on-axis}
	\end{figure}

	In Figure~\ref{fig:on-axis} we present the differential on-axis spectrum of the emitted photons
	$\frac{\ud \mathbb P}{\ud t \ud^2 \vec k_\perp'}|_{\vec k_\perp'=0}$ for electrons which were polarized along the beam axis
	before the scattering, $\Xi_z = 1$.
	We calculate the probability to find an electron with its spin aligned parallel $s'=+1$ (non-flip, $\up$)
	and anti-parallel $s'=-1$ (spin-flip, $\down$) to the $z$-axis by employing the following projection operators,
	\begin{align} \label{eq:onaxis-prob}
	\left. \frac{\ud \mathbb P^{s'}}{\ud t \ud^2 \vec k_\perp'} \right|_{\vec k_\perp'=0} 
				= \tr \left[ \frac{1 + s'\sigma_3}{2} \rho(\vec k_\perp'=0,t)  \right] \,,
	\end{align}
	with $\rho(\vec k_\perp'=0 , t)$ from Eq.~\eqref{eq:rho-differential}.

	The numerical results for the on-axis differential probability
	are presented in Fig.~\ref{fig:on-axis}. They show that the even harmonics 
	(red dashed curves) are much weaker than the two adjacent odd harmonics
	(blue solid curves). The even harmonics are completely due to the spin-flip transitions, while the odd harmonics are produced exclusively in non-flip transitions.
	
	Because all particles are co-axial we can explain this phenomenon by angular momentum conservation.
	Each laser photon and emitted hard photon can be assigned an angular momentum (projection)
	of 1 (unit of $\hbar$). That means for the even harmonics, for instance the second harmonic,
	two laser photons are absorbed by the electron (with an even multiple of angular momentum),
	while the emitted photon carries away only one unit. This imbalance
	is resolved by the electron undergoing a spin-flip transition from spin projection $+1/2$ to $-1/2$.

\begin{figure}[tb]
	\includegraphics[width=0.49\columnwidth]{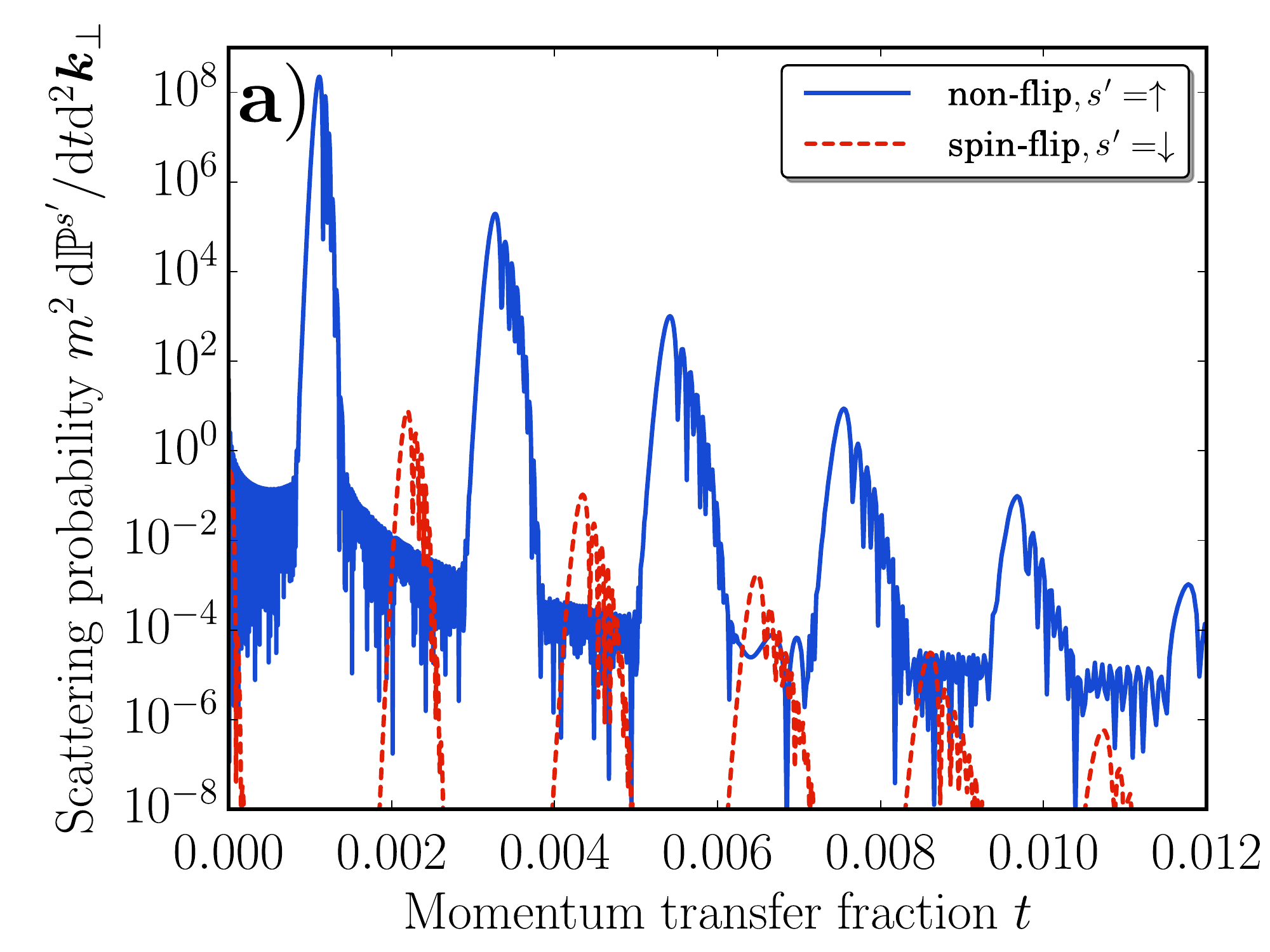}
	\includegraphics[width=0.49\columnwidth]{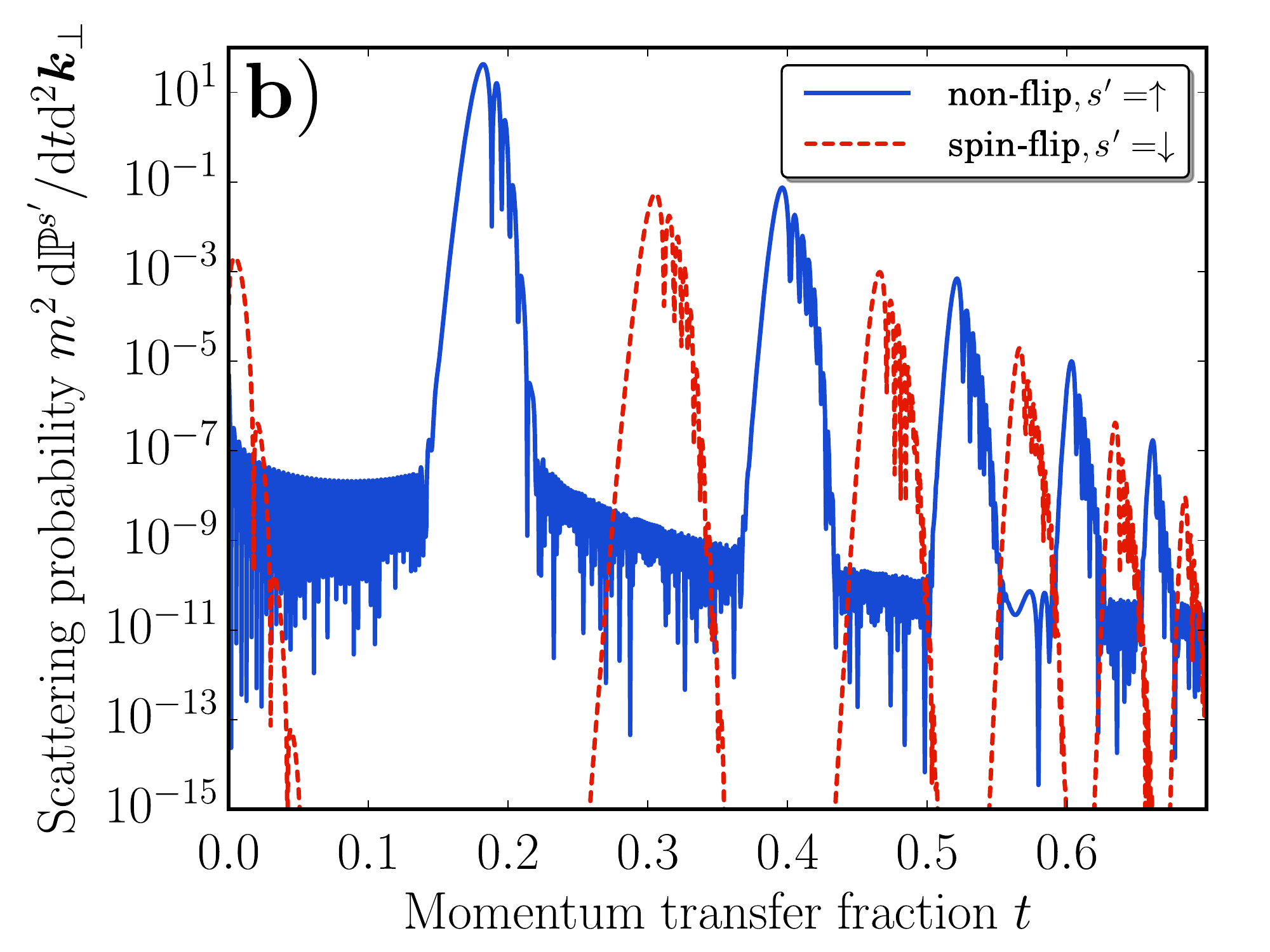}
	\includegraphics[width=0.49\columnwidth]{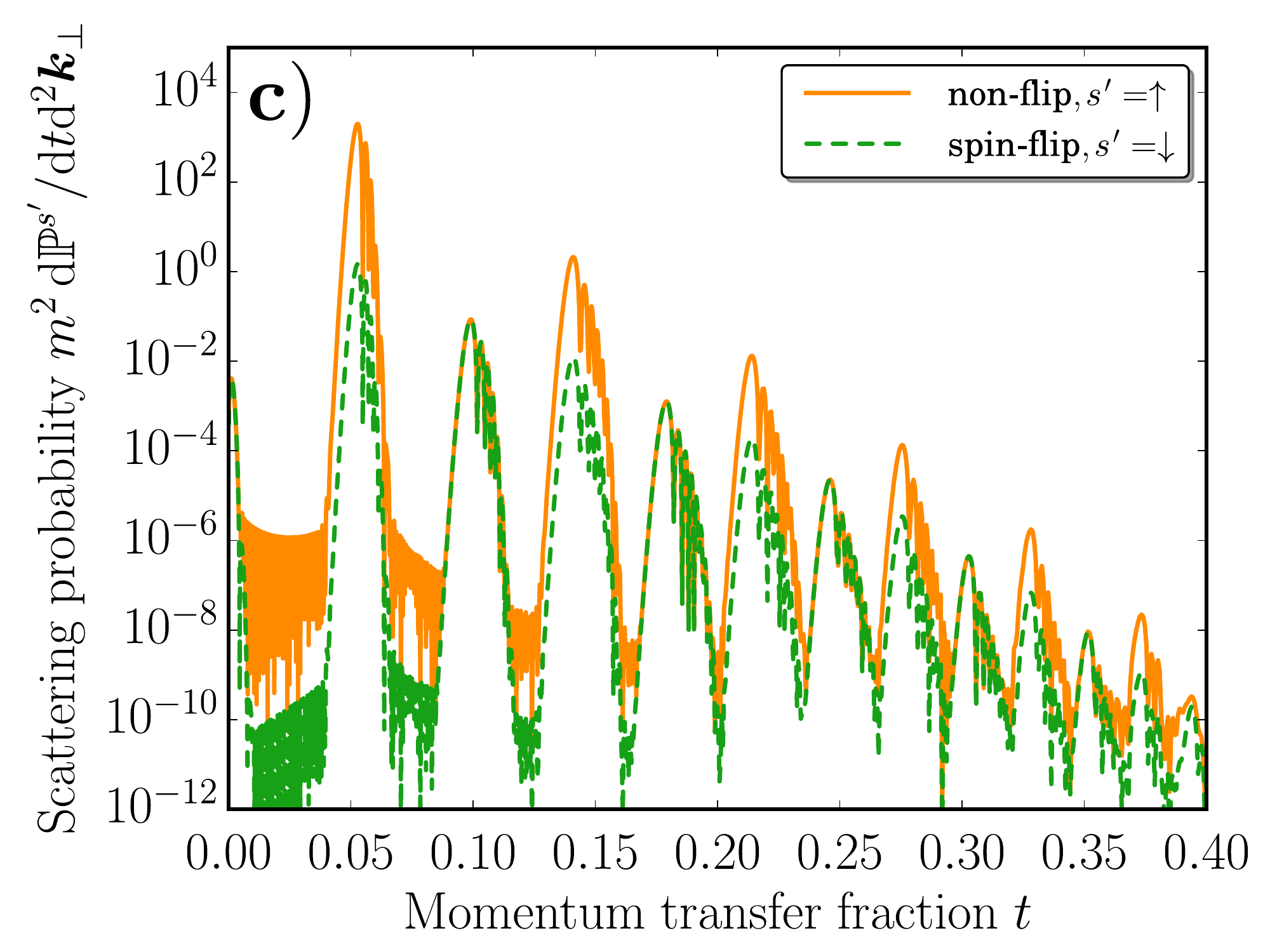}
	\includegraphics[width=0.49\columnwidth]{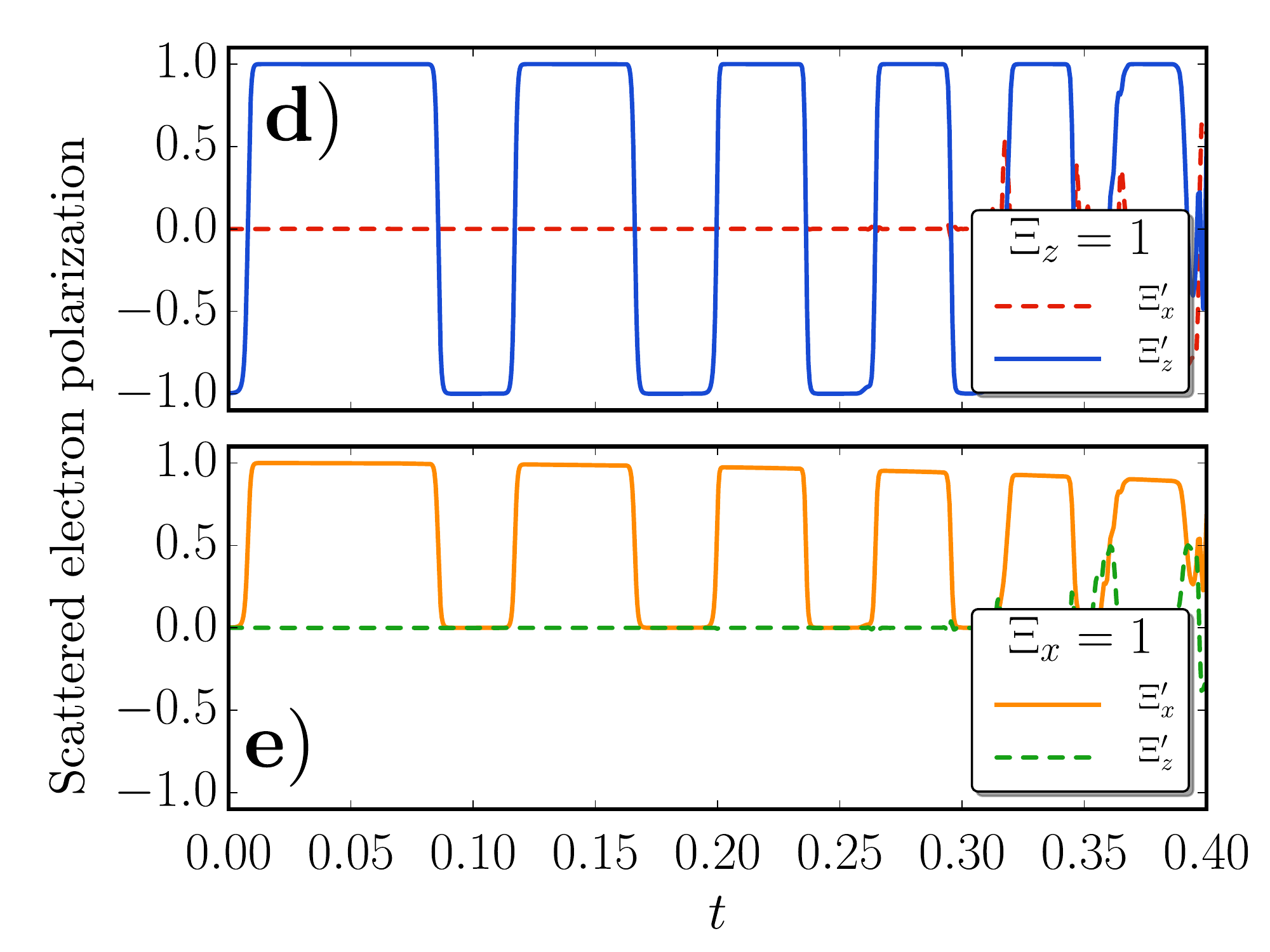}
	\caption{Upper panels: On-axis spectrum for the head-on collision of a polarized electron beam as in Fig.~\ref{fig:on-axis}
	but for $\gamma=100$ (a) and $20000$ (b).
	Lower panels: On-axis spectrum for the head-on collision of an electron beam polarized parallel to the x-axis before the scattering (c). 
	Same parameters as in Fig.~\ref{fig:on-axis}.
	Components of the \Stokes vector of the scattered electrons for electrons initially
	polarized along the $z$-axis (d), or along the $x$-axis (e).} 
	\label{fig:on-axis2}
	\end{figure}

	The suppression of the even harmonics is roughly proportional the quantum energy parameter $b=k.p/m^2$.
	A calculation for smaller and larger initial electron energies of $\gamma=100$ and $20000$ ($b=6\times 10^{-4}$ and $0.12$),
	respectively, presented in Fig.~\ref{fig:on-axis2} (a,b) supports this.
	In particular, that means in the classical low-energy limit $b\ll1$ spin effects become negligible.
	This is consistent with the fact that in classical mechanics and electrodynamics the electron trajectories are decoupled from spin.
	Accordingly, spin transitions are impossible in the classical picture and 
	the even harmonics are not present when the radiation spectra are calculated using the classical Li\'enard-Wiechert potentials
	(nonlinear Thomson scattering) \cite{Esarey:PRE1993,Thomas:PRSTAB2010}.

	What part of the spectrum is formed by spin-flip or non-flip transitions depends strongly
	on the way the electron was polarized before the scattering.
	This is shown in Fig.~\ref{fig:on-axis2} (c) where we plot the spin-flip and non-flip probabilities for electrons that were
	initially polarized along the $x$-axis, i.e.~perpendicular to the beam axis.
	These probabilities are calculated similar to Eq.~\eqref{eq:onaxis-prob} but with $\sigma_3$ replaced by $\sigma_1$,
	i.e. we use the $x$-axis as spin quantization axis.
	In the case of initially $x$-polarized electrons
	the odd harmonics now contain contributions from the spin-flip transitions (green curves), in which the electron is polarized along the
	negative $x$-axis after the scattering, and non-flip transitions (orange curves),
	(compare Fig.~\ref{fig:on-axis2} (c) with \ref{fig:on-axis}).
	Similarly, the even harmonics, which were completely dominated by spin-flip transitions in the $z$-polarized case now contain
	both contributions, and with spin-flip and non-flip transitions contributing equally.

	The polarization of the final electrons depends on the fraction of light-front momentum 
	$t = k'^+/p^+$ transferred from the electron to the photon, and on the initial polarization of the electron.
	For instance, if the electron was polarized along the $z$ direction before the scattering, it will be completely polarized after the scattering, but
	with its direction depending on $t$.
	This can be seen in Fig.~\ref{fig:on-axis2} (d), where the final electron \Stokes vector component $\Xi_z'(t)$ flips
	between $+1$ and $-1$ depending on $t$, while $\Xi_x' =0$.
	Contrary, if the electron was polarized along the $x$-axis before the scattering
	(Fig.~\ref{fig:on-axis2} e) then it will be unpolarized for certain values of $t$ after the scattering,
	where both $\Xi_z' = \Xi_x'=0$. The component $\Xi_y'$ vanishes in both cases.

	What these examples show is that one has to be very careful when talking about spin-flip transitions and non-flip transitions
	since the corresponding rates and the final electron polarization depend strongly on the orientation of the electron spin before the scattering.
	If one does not employ the density matrix formalism the initial polarization direction coincides with the chosen basis for the quantization of the
	spin.
	In fact, there was some discussion in the literature whether or not spin-flip transitions are relevant and the electrons
	do spin-polarize \cite{Bagrov:INCB1989,Kotkin:PRSTAB2003,Karlovets:PRA2011}.
	These discrepancies were attributed to different choices of the quantization direction \cite{Karlovets:PRA2011}.
	We stress here that no such ambiguities arise when working in the density matrix formalism, where the final electron
	\Stokes vector unambiguously describes the polarization after the scattering. It's value is an observable and
	independent of the choice of the quantization axis for the spin. 
	We will return to the problem of spin-flip transitions later when we calculate analytically the spin-flip transition probabilities
	for an arbitrary direction of the initial electron spin polarization in Section \ref{sect:flip-transitions}.

	\subsection{Electron Polarization of Short Circular Laser Pulses}

	We now investigate the electron polarization after they emitted a Compton photon when interacting with
	a short circularly polarized laser pulse,
	$A^\mu =  m\xi \cos^2 (\frac{ \pi\phi}{2 \Delta\phi}) \, \Theta( \Delta\phi - | \phi | )  (0,\cos\phi,\sin\phi,0)/\sqrt{2}$,
	where $\Theta$ is the Heaviside step function.

\begin{figure}
	\includegraphics[width = 0.90\columnwidth]{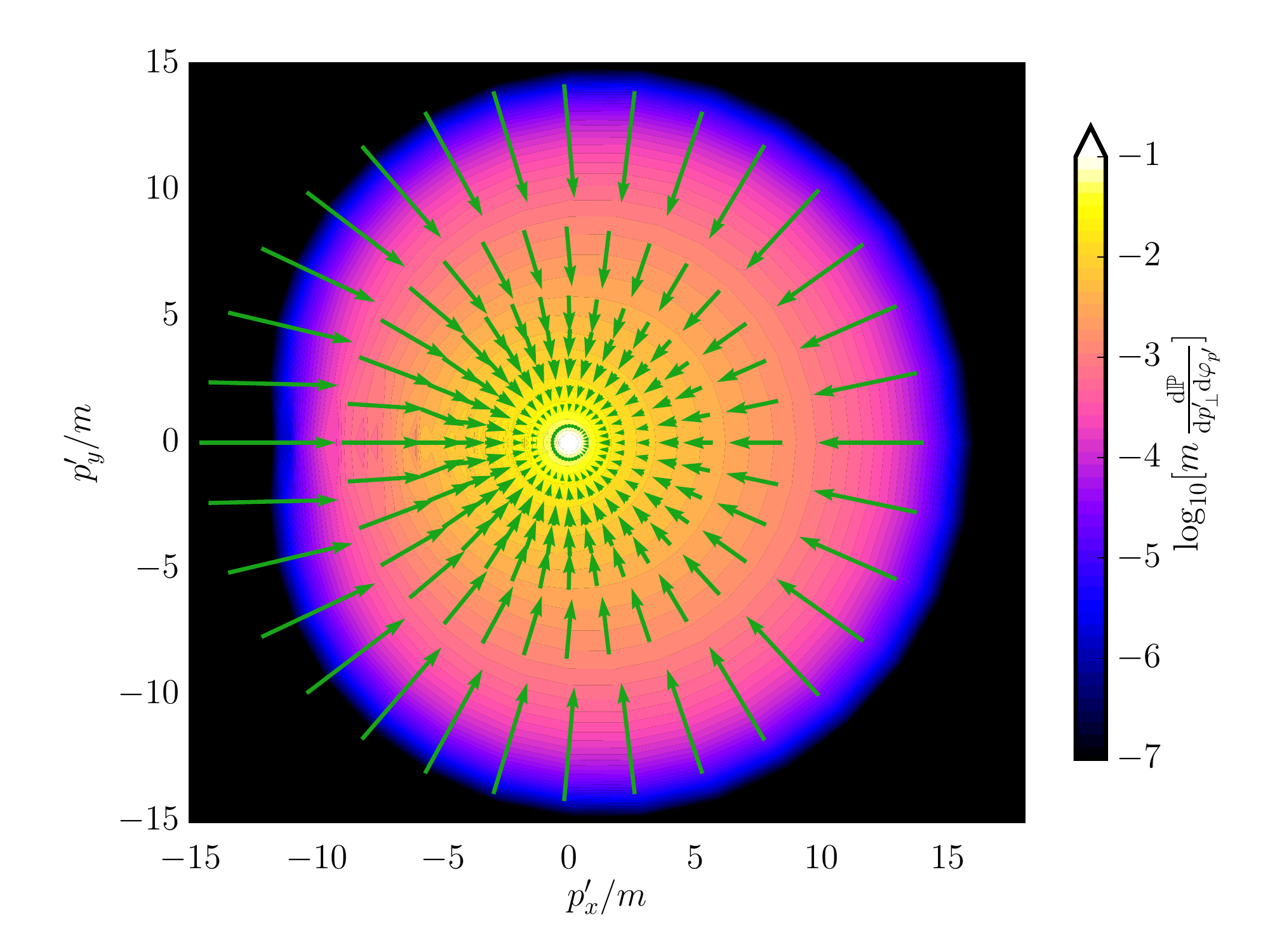}
	\caption{Transverse momentum distribution of the scattered electrons (as a heatmap)
	and the polarization of the scattered electrons transverse to the beam axis (arrows). The length of the arrows indicates
	the magnitude of the polarization for a given $\vec p_\perp'$. Parameters are $\gamma=5000$, $\xi=25$, and $\Delta \phi = 3\pi$.}
	\label{fig:beam-profile}
	\end{figure}

	The transverse differential photon emission probability $\ud \mathbb P/\ud  p_\perp' \ud \varphi_{p'}$ 
	(integrated over the longitudinal light-front 	momentum fraction $t$) of the
	electrons after the scattering is shown in Figure \ref{fig:beam-profile}, for
	$\gamma=5000$, $\xi=25$, and $\Delta \phi = 3\pi$.
	The azimuthal symmetry that one might naively expect of the emission probability to hold for a circularly polarized pulse
	is broken for an ultra-short short laser pulse. This behavior is in line with what is known form the literature \cite{Seipt:PRA2013,Titov:PRD2016}.

	The direction of the electron polarization in the transverse $x$-$y$ plane
	is indicated by the green arrows in Fig.~\ref{fig:beam-profile},
	where the length of the arrows is proportional to the degree of polarization.
	Electrons that are scattered to larger transverse momenta $p_\perp'$ are more strongly polarized,
	there is a strong correlation between the momenta of the scattered electrons and their polarization.
	The direction of the transverse polarization seems to be almost radial.
	But again, as for the probability, the azimuthal symmetry is broken due to the short duration of the considered few cycle laser pulse.
	This azimuthal symmetry breaking becomes less pronounced for longer pulses.

	Because the polarization pattern is almost radial, it is quite useful to 
	describe the polarization here by means of the components of the final electron \Stokes vector with regard to the scattering plane.
	The latter is spanned by the directions of the incident and scattered electron momenta $\hat{\vec p} = \vec p /|\vec p|$
	and $\hat{\vec p}' = \vec p' /|\vec p'|$, where the hat indicates a unit vector.
	In addition to the longitudinal polarization vector  $\vec q_{\parallel}  = \hat{\vec p}'$
	we define the two polarization vectors perpendicular to $\vec p'$,
	$\vec q_{\perp,\rm out} 
			= (\vec q_{\parallel} \times \hat{\vec p} )/
						 | \vec q_{\parallel} \times \hat{\vec p} | $, and
	$\vec q_{\perp,\rm in} 
			= ( \vec q_{\perp,\rm out} \times \vec q_{\parallel} )/
						{ | \vec q_{\perp,\rm out}\times \vec q_{\parallel}  |} $.
	The vectors $\vec q_\parallel$ and $\vec q_{\perp,\rm in} $ are lying in the scattering plane,
	while $\vec q_{\perp,\rm out} $ stands perpendicular to the scattering plane. Any non-zero value of
	the electron polarization out of the scattering plane, $\vec \Xi'  \cdot \vec q_{\perp,\rm out} $, indicates a breaking the azimuthal symmetry.
	While for a single-cycle pulse, $\Delta \phi = 2\pi$, the maximum of the out-of-plane
	polarization reaches $0.002$ for $\xi = 14.1$ and $\gamma=1000$,
	it is smaller than $5\times 10^{-5}$ for a two-cycle pulse, $\Delta \phi = 4\pi$, and otherwise same parameters.

\begin{figure}[bt]
	\includegraphics[width=\columnwidth]{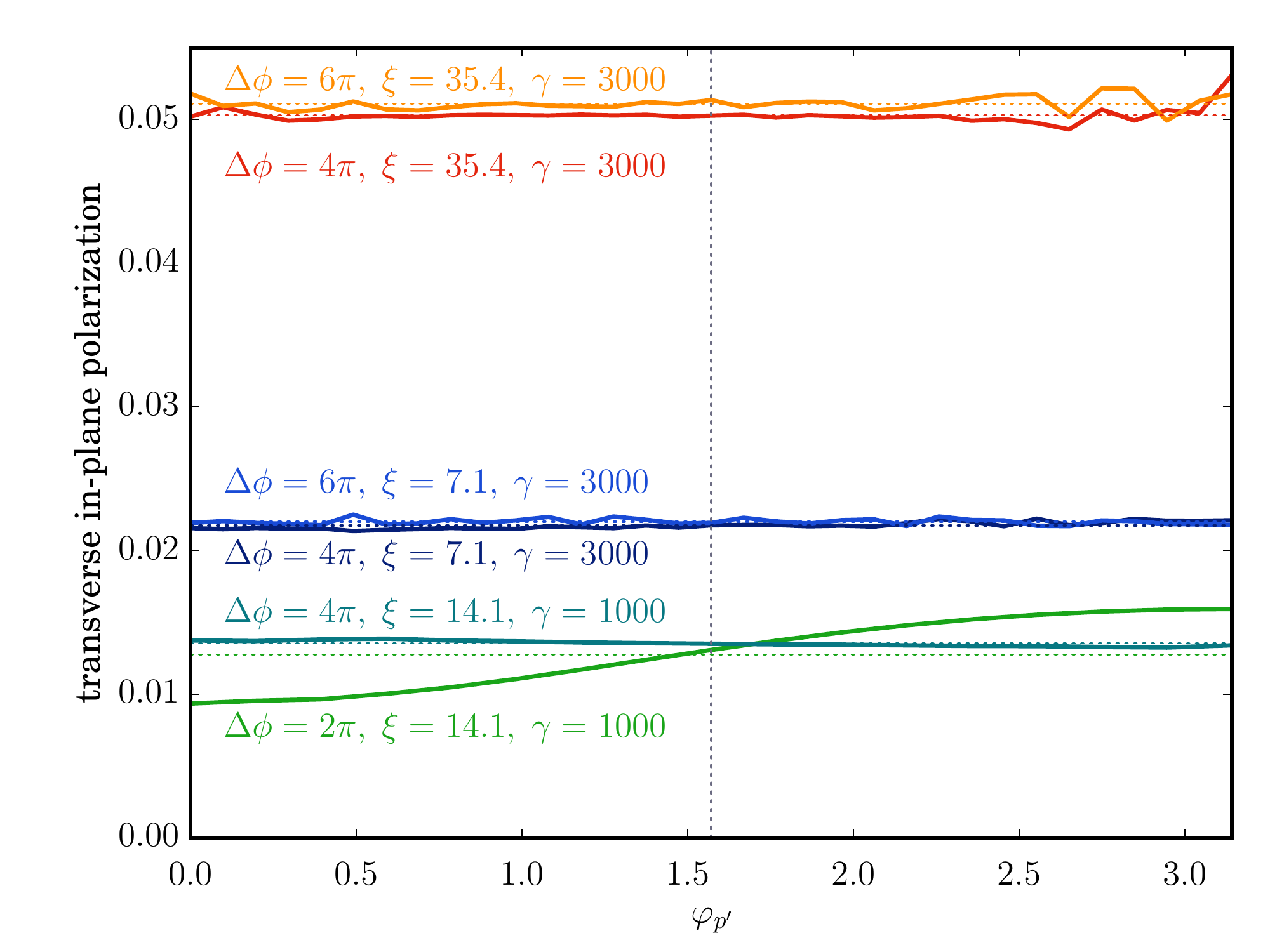}
	\caption{Degree of transverse polarization in the scattering plane as a function of azimuthal angle
	for various combinations of laser intensity $\xi$, pulse duration $\Delta \phi$ and initial electron energy $\gamma$.
	The dashed lines correspond to the averaged values over all angles.}
	\label{fig:angular}
	\end{figure}

	Furthermore, we see also strong azimuthal asymmetries in the in-plane transverse polarization
	for ultra-short pulses, see in Figure \ref{fig:angular}.
	The shorter single-cycle laser pulse, $\Delta \phi = 2\pi$, (green curve)
	shows a much more pronounced azimuthal dependence of the electron polarization
	as compared to the longer pulses.
	The values at $\varphi_{p'} = \pi/2$, i.e.~perpendicular to the direction of the peak of the vector potential, are in good agreement with
	the mean values averaged over all angles (dashed lines).

	\subsection{Dependence of the Polarization Degree on $\xi$ and $\chi_e$}

\begin{figure}[!ht]
	\includegraphics[width=0.99\columnwidth]{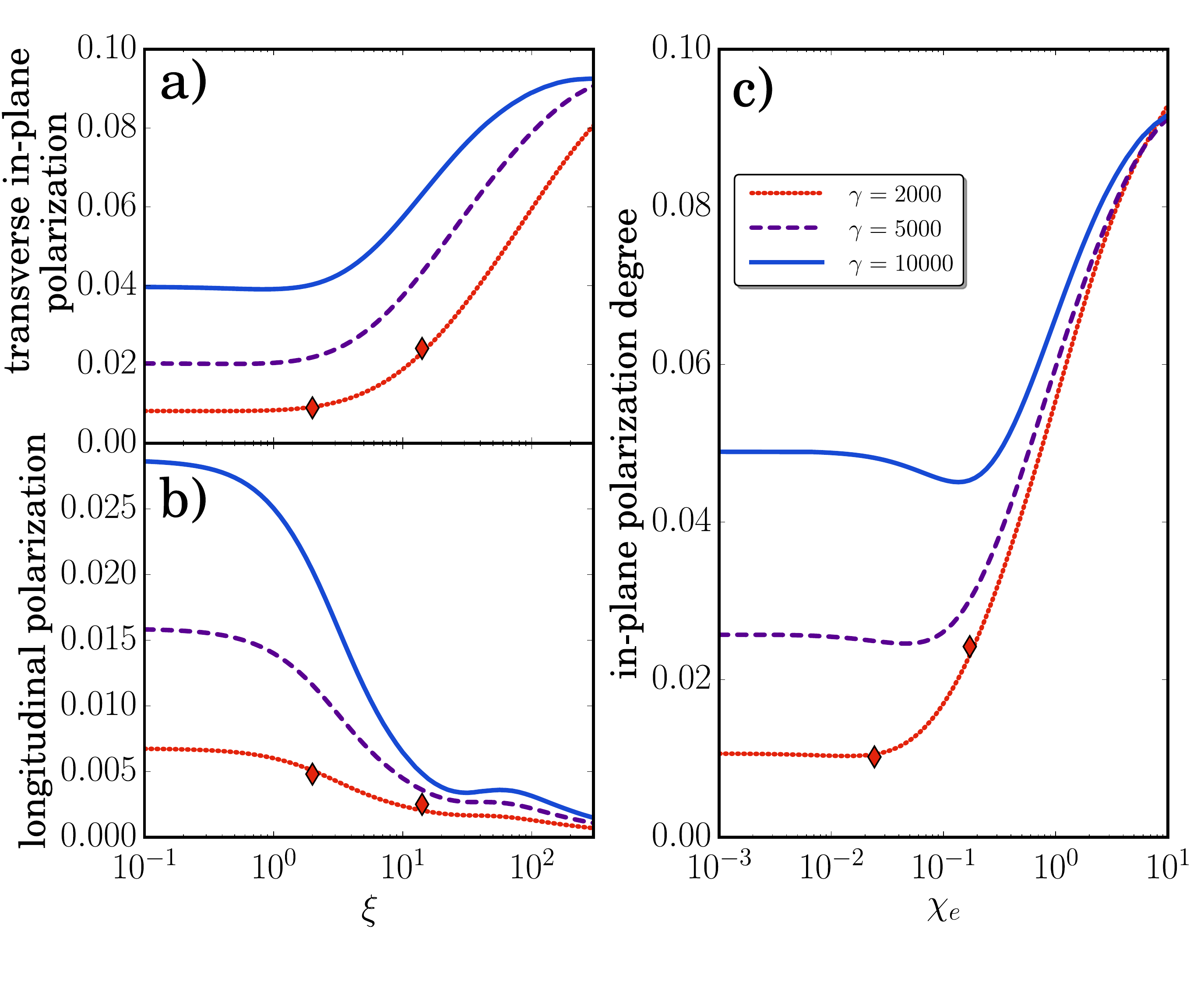}
	\caption{Polarization in the scattering plane as a function of laser intensity.
	(a) Degree of transverse in-plane polarization as a function of $\xi$;
	(a) degree of longitudinal polarization as a function of $\xi$;
	(c) degree of polarization in the scattering plane as a function of $\chi_e$. 
	The symbols are integrals over the complete phase space (all azimuthal angles).
	}
	\label{fig:scan}
	\end{figure}

	In this section we investigate how the degree of polarization depends on the laser intensity parameter $\xi$
	and the quantum efficiency parameter $\chi_e$.
	The results are shown in Fig.~\ref{fig:scan}. In (a) we plot the transverse polarization in the scattering plane as a function of
	laser strength $\xi$ for three different initial electron energies (red,blue and purple curves),
	and for electrons emitted at $\varphi_{p^\prime} = \pi/2$.
	In the low-intensity region $\xi\ll1$ the polarization degree is independent of the laser intensity, and
	larger for larger values of $\gamma$, i.e. for larger quantum energy parameter $b\approx 2\gamma \omega/m$.
	For instance, for $\gamma = 10000$ ($b = 0.06$) the low-intensity transverse polarization is $0.04$.
	With increasing $\xi$ the transverse polarization in the scattering plane increases, reaching values of about $0.09$ at $\xi=300$.

	The longitudinal polarization (Fig.~\ref{fig:scan} (b)) shows an opposite behavior:
	Its maximum values are achieved in the low-intensity region, and they drop for increasing $\xi$.

	When viewed as a function of the quantum efficiency parameter $\chi_e = \xi b$, the degree of polarization in the scattering plane
	(Fig.~\ref{fig:scan}) is independent of $\chi_e$ in the semiclassical limit for small $\chi_e\ll1 $,
	but its value depends on $\gamma$.
	For large $\chi_e$ the degree of polarization increases up to $0.09$ for 
	$\chi_e=10$, with all three curves for different $\gamma$ showing a similar behavior.

	The degree of polarization for fixed azimuthal angle $\varphi_{p^\prime} = \pi/2$ serves as a reliable
	representative for the polarization degrees averaged over all $\varphi_{p^\prime}$, even for very short pulses as discussed above.
	For longer pulses, where the azimuthal symmetry is restored, they coincide exactly.
	To validate the agreement between the two we compare the fixed angle polarization
	with a few calculations where we integrate over the full phase space of the outgoing electron, including
	the azimuthal angle. These latter results are the diamond symbols in Fig.~\ref{fig:scan}.

	\section{The Locally Constant Crossed Field Approximation for the Polarization Density Matrix}
	\label{sect:analytic}

	The numerical investigations in the previous section have shown that a significant degree of polarization can be
	achieved already after emission of a single photon in an ultra-short laser pulse.
	However, in strong fields and for longer pulses multi-photon emission has to be taken into account. In fact, in a strong field the
	radiation length, i.e. the mean phase distance between between two photon emissions, scales as
	$ L_C \sim 95 /\xi$ for $\chi_e\ll1$ \cite{Bulanov:PRA2013}.
 	By calculating strong-field QED S-matrix elements one connects initial and final asymptotic states.
	A proper inclusion of multi-photon emission effects in the QED framework would require to
	evaluate the S-matrix elements for higher-order Feynman diagrams with more than one photon in the final state.
	In the presence of a strong laser field
	such calculations have been performed up to $n=2$ emitted photons only (nonlinear double Compton scattering)
	\cite{Lotstedt:PRL2009,Seipt:PRD2012,King:PRA2015}, and the calculations are numerically demanding.
	The influence of electron spin has been explicitly addressed in Ref.~\cite{King:PRA2015} for scattering in a constant crossed field.
	Explicitly time-dependent Hamiltonian approaches to strong-field QED have been investigated,
	but they pose additional difficulties \cite{Ilderton:PRD2013b,Dinu:PRL2016}.

	The standard way of simulating such multi-photon emissions relies on the fact that the photon {formation phase interval}
	becomes short at large field strength, $\propto 1/\xi$.
    The multi-photon emission process is assumed to factor into
	a sequence of one-photon emission events with classical propagation between the emission vertices.
	Photon emission is usually described via Monte Carlo sampling from the (spin and polarization averaged) one-photon emission rates
	$\ud \mathbb P/\ud t \ud \phi$ in a locally constant crossed field
	\cite{Ridgers:JCompPhys2014,Gonoskov:PRE2015,Green:CPC2015,Harvey:PRA2015,Blackburn:2018}.

	To be able to study radiative electron polarization effects in this framework we derive in this section
	analytically the locally constant crossed field approximation of the polarization density matrix as
	generalization of the known LCFA photon emission rates \cite{book:Sokolov,book:Baier,Bagrov:INCB1989,King:PRA2013}.
	We will focus here on the case of linear laser polarization, where the laser vector potential 
	$A^\mu = m \xi \varepsilon^\mu h(\phi)$ is determined by the normalized shape function $h(\phi)$.
	Moreover, we use the special physical spin-basis introduced in Section \ref{sect:basis} from now on.
	
	To calculate the LCFA expressions for the polarization density matrix
	 we first need to analytically evaluate the Dirac spinor structures in \eqref{eq:rho-differential}.
	They can be expressed as traces over Dirac indices by replacing the products of the spinors with help of \eqref{eq:uubar},
	\begin{multline} \label{eq:general-trace}
	(\bar u_{p' \bar  s'} \mathscr J^\mu(\phi) u_{ps}) \: ( \bar u_{p \bar s} \bar{\mathscr J}_\mu(\phi')  u_{p's'}) 
	=
	 \frac{1}{4} \tr_\mathrm{D} \bigg[ (\slashed p' + m)  \\
	\times ( \delta_{s'\bar s'} - \gamma^5 \slashed S'_{s'\bar s'} ) 
	 \mathscr J^\mu(\phi) 
	 (\slashed p + m ) ( \delta_{s\bar s} - \gamma^5 \slashed S_{s \bar s}  ) \bar{\mathscr J}_\mu(\phi')
	\bigg]
	\end{multline}
	which can be evaluated using standard techniques \cite{book:Itzykson}.
	All Dirac traces can be expressed in terms of four basic trace expressions, which we call $\mathsf{UP}$ (standing for unpolarized electrons),
	$\mathsf{IP}$/$\mathsf{FP}$ (referring to initially/finally polarized electrons) and $\mathsf{PC}$ (describing effects of polarization
	correlation between initial and final electrons).
	Explicit expressions for those basic traces are collected in Appendix \ref{sect:traces}.

	After having performed all the Dirac traces we obtain
	the \textit{reduced polarization density matrix} in the following form:
	\begin{widetext}
	\begin{multline} \label{eq:final-density-matrix-allintegrals}
	\rho
			= - \frac{\alpha\xi^2 }{16\pi^2 m^4 \chi_e^2 }
				\int \frac{\d t }{ t (1-t) } \, \d^2\vec k'_\perp \, 
				\d\sigma \d\tau   \:  e^{i \sigma \frac{k'.\langle \pi \rangle}{k.p'}} \\
			  \times 
				\left(
			\begin{matrix}
				\mathsf{ UP } - \mathsf{FP}(\zeta') 	- \mathsf{IP} ( \Xi )		+ \mathsf{PC} (\zeta' , \Xi ) &
				- \mathsf{FP}(e'_-) +  \mathsf{PC}(e'_- , \Xi  )  \\
				 - \mathsf{FP}(e'_+) +  \mathsf{PC}(e'_+, \Xi ) &
				  \mathsf{ UP } + \mathsf{FP}(\zeta') 	- \mathsf{IP} ( \Xi ) - \mathsf{PC} (\zeta' , \Xi )
				  \end{matrix}
				\right)
				\,.
	\end{multline}
	\end{widetext}
	Several terms in \eqref{eq:final-density-matrix-allintegrals} depend on
	the \Stokes four-vector $\Xi^\mu = \Lambda^{\mu}_{\ \nu}(\vec p) (0,\vec\Xi)^\nu$,
	which is the initial electron \Stokes $\vec \Xi$ vector boosted from the electron rest frame to the laboratory frame.
	In addition, we transitioned from the phase variables $\phi',\phi$ to their midpoint
	$\tau = (\phi + \phi')/2$ and relative phase $\sigma = \phi'-\phi$
	and introduced the \floating average between $\phi$ and $\phi'$,
	\begin{align}
	\langle \pi \rangle  = \frac{1}{\sigma} \intop_{-\sigma/2}^{\sigma/2 } \! \ud \theta  \: \pi( \theta +\tau ) \,,
	\end{align}
	i.e. over the interval $\sigma$ with midpoint $\tau$.

	Before discussing the spin-polarization of the scattered electrons any further let us first investigate the total scattering probability
	by calculating the trace $\mathbb P = \tr \rho$, with $\rho$ from Eq.~\eqref{eq:final-density-matrix-allintegrals}.
	In the case of unpolarized initial electrons, $\Xi^\mu = 0$, and by using expression \eqref{eq:UP-trace} for $\mathsf{UP}$
	we immediately find
	\begin{multline}
	\mathbb P	= - \frac{ \alpha \xi^2  }{4\pi^2 m^2 \chi_e^2  } 
			\int \frac{ \ud t }{  t(1-t) } \: \d^2\vec k'_\perp   
            \: \d\sigma \d\tau  \:  e^{i \sigma \frac{k'.\langle \pi \rangle}{k.p'}}  \\
			\times \left( 1 + \frac{\xi^2 g }{2} \sigma^2 \langle \dot h \rangle^2  \right) \,,
	\end{multline}
	with $t = \chi_\gamma/\chi_e = k.k'/k.p$, and
	which agrees with the literature \cite{Dinu:PRA2013}.
	Moreover, it was shown in \cite{Dinu:PRA2013} that the integrals over the transverse photon momenta $\ud^2 \vec k_\perp'$ are
	Gaussian and can be readily performed. We shall now generalize this to the polarized parts.

	\subsection{Local Constant Field Approximation}

	For the considered parameters $\xi\gg1$, 	$p^+\gg m\xi$ and $p_\perp\ll p^+$ the Compton photons are emitted in a narrow cone of opening
	angle $1/\gamma$ around the direction of the electron. 
	As a reasonably good approximation, this angular distribution of the emitted photons is 
	usually not considered, the photons are generated with a momentum parallel
	to their parent electron momentum, and the recoil momentum for the electron is approximated accordingly \cite{Ridgers:JCompPhys2014}.
	This is what is done customarily in
	QED laser-plasma simulation codes \cite{Ridgers:JCompPhys2014,Gonoskov:PRE2015},
	where the (unpolarized) transverse momentum integrated LCFA photon emission rates are used for
	Monte Carlo sampling of the emitted photon spectrum.
	Note that the final electron polarization basis vectors, Eqs.~\eqref{eq:spin-basis-lab} and \eqref{eq:spin-basis-lab-add},
	then are calculated using the approximated final electron momentum vectors.

	Analogously, we calculate here the transverse momentum integrated reduced polarization density matrix of the scattered electrons
	as a first step towards the LCFA. The details of the calculation can be found in Appendix \ref{sect:transverse:integral},
	and the result reads
	\begin{align} \label{eq:final-density-matrix-kperp-done}
	\rho
	=  \frac{\alpha }{4 \pi b }
			\int \d t   \,  \d\tau \, \frac{ \d\sigma }{ i \sigma }
	 e^{ i \sigma \frac{   \mathscr M^2  }{2 b m^2} \frac{t}{1-t} } 
        \left(
				\begin{matrix}
				\mathcal V_+ &
						 \mathcal T^* \\
			            \mathcal T  &
				 \mathcal V_-
				  \end{matrix}
				\right) 
				\,,
	\end{align}
	with Kibble's effective mass $\mathscr M^2$, Eq.~\eqref{eq:kibble:mass},
	and where we introduced the short-hand notations
	\begin{align}
	\label{eq:V}
		\mathcal V_\pm 
			& = 
					1 \pm \Xi_\zeta + \sigma^2 \frac{\xi^2  \langle \dot h \rangle^2}{2} ( g \pm \Xi_\zeta ) \nonumber \\
				& \qquad \qquad  -  i \sigma \frac{\xi}{2}  \langle \dot h \rangle \left[ \Xi_\zeta t \pm \frac{t}{1-t}  
                \right]		\,, \\
	\mathcal T 
				& = 
					\Xi_\kappa \left[  2g - 1 + \frac{g}{2}  \xi^2 \sigma^2 \langle \dot h \rangle^2 \right] 
                    + i \Xi_\eta \left[ 1 + \frac{\xi^2\sigma^2 \langle \dot h\rangle^2}{2} \right] \nonumber \\
				& \qquad \qquad   + i \xi \delta h \left[ \frac{t}{1-t} \Xi_\kappa +it \Xi_\eta \right] \,, 
      \label{eq:T} 
	\end{align}
	where $g$ is defined in Eq.~\eqref{eq:def-g}.
	We see that the off diagonal elements of the density matrix depend only on $\Xi_\kappa$ and $\Xi_\eta$,
    but not $\Xi_\zeta$, while the diagonal elements depend on $\Xi_\zeta$ only.
    In other words, the diagonal elements of the final electron polarization density matrix are only sensitive to
    the initial electron polarization along the direction of the magnetic field (in the initial particle rest frame).
    This is a {great} simplification compared to Eq.~\eqref{eq:final-density-matrix-allintegrals}.

	The locally constant field approximation (LCFA) relies on the fact that one can neglect the spatio-temporal variation of the
	background field in the region of the formation of the process. 
	The latter can be described by the {formation interval} of the emitted photon, which becomes very short $\propto 1/\xi$
	for $\xi\gg1$. 
	Note, however, that this behavior of the {formation interval} depends also on the energy of the emitted photon
	and is strictly valid only for high-energy photons.
	For a proper description of the low-frequency part of the spectrum, coherences over large distances are relevant
	\cite{Harvey:PRA2015,DiPiazza:PRA2018}. This affects both shape of the spectrum for low-energy photons,
	as well as the integrable IR-divergence $\propto t^{-2/3}$ in the LCFA which is absent in the exact result.
	Moreover, for radiation reaction effects, i.e.~the backreaction of photon emission on the electron,
	the low-frequency part of the spectrum is not important.	
	
	In the LCFA, the laser field can be considered as constant over the
    {formation interval},
	and interferences outside the {formation interval} are neglected because they are unimportant.
	To calculate the LCFA for the density matrix of the scattered electrons we approximate the
	\floating averages $\langle h \rangle = \sigma^{-1} \int_{\tau-\sigma/2}^{\tau+\sigma/2} 
	\! \ud \phi \:  h(\phi)$ for short intervals $\sigma \ll1$.
	For instance, we approximate the \floating averages in the pre-exponential in Eqs.~\eqref{eq:T} and \eqref{eq:V}
	to lowest order as $\langle \dot h \rangle \simeq \dot h(\tau)$,
	because for very short averaging intervals the \floating average approaches the value at the midpoint.
	In addition, in the off-diagonal terms we may use $\delta h \simeq 0$ for locally constant fields, see Appendix \ref{sect:transverse:integral}.

	In order to expand Kibble's mass $\mathscr M^2$ appearing in the exponent in \eqref{eq:final-density-matrix-kperp-done}
	we need to  go up to second order in $\sigma$,
	$$\mathscr M^2 \simeq m^2  +   \frac{1}{12} m^2 \xi^2 \sigma^2 \dot h^2(\tau)  \,,$$
	so that the exponent is approximated as
	\begin{align}
	 e^{ i \sigma \frac{ \xi  \mathscr M^2  }{2 \chi_e m^2} \frac{t}{1-t} }  
	\simeq
	 e^{ i \frac{ \xi  }{2 \chi_e } \frac{t}{1-t} \left( \sigma +  \frac{\sigma^3}{12} \xi^2 \dot h^2 \right)   } 	
	 = e^{i x \sigma + i \frac{y}{3}\sigma^3 }
	\end{align}
	with 
	\begin{align}
	x = \frac{\xi t}{2\chi_e(1-t)} \,, \qquad
	y = \frac{x \xi^2}{4} \: \dot h^2  \,, \qquad
	z = \frac{x}{\sqrt[3]{y}} \,.
	\end{align}

	With these approximations, we can perform the integrals over $\ud \sigma$
	in \eqref{eq:final-density-matrix-kperp-done}, yielding Airy functions,
	$\Ai(z) = \frac{1}{2\pi} \int_{-\infty}^\infty \! \ud s \: e^{i\frac{s^3}{3}+izs}$,
	their derivative $\Ai'(z)$, and integral $\Ai_1(z) = \int_z^\infty \! \ud x \: \Ai(x)$.

	As our final analytical result for the polarization density matrix of the final electrons in the local constant crossed field approximation
	we obtain the following expressions:
	\begin{align} \label{eq:rho-LCFA-final}
	\rho
		= - \frac{\alpha }{2b} \! \intop \! \ud \tau \ud t \: 
	\left( 
	\begin{matrix}
	\tilde {\mathcal V}_+ & \tilde{\mathcal T}^* \\ \tilde{\mathcal T} & \tilde{\mathcal V}_- 
	\end{matrix}		
	\right)
	\,,
	\end{align}
	with
	\begin{align}
	\label{eq:V-LCFA}
	\tilde{\mathcal V}_\pm & =
			( 1  \pm \Xi_\zeta ) \Ai_1(z)  
		+  (g \pm \Xi_\zeta) \, \frac{2 \Ai'(z) }{z} \nonumber \\
		&\qquad +  \left(  t \Xi_\zeta \pm  \frac{t}{1-t}  \right)  {\rm sign}(\dot h)\:  \frac{ \Ai(z) }{\sqrt{z}}	
		\,,\\
		\label{eq:T-LCFA}
		\tilde{\mathcal T} &= 
				\Xi_\kappa \left[ 
								(2g-1)  \Ai_1(z) +  g \, \frac{2 \Ai'(z)}{z}  
								\right] \nonumber \\
		&\qquad	+ i \Xi_\eta \left[  \Ai_{1}(z) + \frac{2\Ai'(z)}{z}  \right] \,,
	\end{align}
	with $g$ from \eqref{eq:def-g},	and $t = k.k' / k.p$ as the light-front momentum fraction transferred from the initial electron to
	the emitted photon.
	The argument of the Airy functions, $z = z(t,\tau) = [\frac{t}{1-t} \frac{1}{| \dot h(\tau) | \chi_e }]^{2/3}$, depends on the 	
	\textit{local}
	value $\chi_e(\tau) \equiv \chi_e | \dot h (\tau) |$ of the quantum efficiency parameter, which is calculated using the
	\textit{local} value of the background field.
	Note that $\dot h$ refers to the shape of the electric field since $h$ is the shape of the vector potential.
	
	This expression \eqref{eq:rho-LCFA-final} for the density matrix, together with \eqref{eq:V-LCFA} and \eqref{eq:T-LCFA}
	contains the complete information on the electron polarization after the emission of a Compton photon, for arbitrary initial electron polarization.

	\subsection{Results and Discussion}

	Before discussing the new results for the polarized parts of the density matrix
	let us first compare with known expressions for unpolarized electrons.
	Taking the trace of \eqref{eq:rho-LCFA-final} and setting the initial \Stokes parameters to zero, the obtained 	
	expressions for the \textit{probability rate} for non-linear Compton scattering,
	\begin{align}
	\mathbb R  
    & = \frac{\ud \mathbb P}{\ud \tau}= \tr [\rho] \nonumber \\
	&= - \frac{\alpha }{b} \! \intop \! \ud t \: \left[ \Ai_1(z) + g \frac{2 \Ai'(z)}{z} 
	  +  \Xi_\zeta  {\rm sign}(\dot h)\:  t \frac{ \Ai(z) }{\sqrt{z}}	 \right] \,,
	\end{align}
	depends only on $\Xi_\zeta$. Moreover, for unpolarized electrons, $\Xi_\zeta=0$, the expression
	completely agrees with corresponding expressions from the literature,
	e.g.~Refs.~\cite{Ritus:JSLR1985,Elkina:PRSTAB2011}.
			
	By taking traces of \eqref{eq:rho-LCFA-final} with different projection operators we can easily determine the probabilities
	to measure the electron with a certain polarization after the scattering. For instance,
	\begin{align}
	\mathbb P_{\zeta'}(s')  = \TR{  \rho \frac{1 + s' \sigma_3}{2}} 	= 	- \frac{\alpha  }{2b} \! \intop \! \ud \tau \ud t \: 
	\tilde {\mathcal V}_{s'} \,,
	\end{align}
	with $\tilde {\mathcal V}_{s'}$ from Eq.~\eqref{eq:V-LCFA},
	gives the probability to find the electron spin aligned or anti-aligned,
    $s' = \pm1$, with the direction $\zeta'$, i.e. polarized parallel
	or anti-parallel to the magnetic field in the rest frame of the scattered electron.
    As we see from Eq.~\eqref{eq:V-LCFA}, these probabilities
	do not depend on the components of the initial electron \Stokes vector other than $\Xi_\zeta$.

	Similarly, the probability to measure the electron polarization (anti-)parallel to the electric field in the electrons rest frame
	is given by
	\begin{align}
	\mathbb P_{\eta'}(s') &  = \TR{  \rho \frac{1 + s' \sigma_2}{2}}
        \nonumber \\
	 	&= \frac{\mathbb P}{2} - s' \Xi_\eta \: \frac{\alpha  }{2b} 
        \! \intop \! \ud \tau \ud t \: \left[ \Ai_1(z) + \frac{2\Ai'}{z} \right] \,.
	\end{align}
	For unpolarized initial electrons, $\Xi=0$, this simplifies to $\mathbb P_{\eta'}(s') = \mathbb P/2$. 
    That means, for electrons that were not polarized along $\eta$ initially
	do not polarize along the $\eta'$ direction. (Similar expressions can be found for the polarization along $\kappa'$.)
	However, electrons polarized along the $\eta$ direction initially
    will lose some degree of this polarization when emitting a photon.	
	The leading $\mathbb P/2$ is a reflection of the fact that,
	for unpolarized electrons, the probability to measure the spin parallel or antiparallel to any direction is $1/2$ each.

	The explicit expressions for the components of the final electron \Stokes vector read
	\begin{align}
	\Xi_{\zeta'} 
			 & = - \frac{\alpha}{b \mathbb P} \int \! \ud \tau \ud t \: \left [ 
			  \Xi_\zeta \Ai_1 + \Xi_\zeta \frac{2\Ai'}{z} \right.  \nonumber \\
			& \qquad \qquad  \qquad\qquad \qquad \left.  + \frac{t}{1-t} \frac{\Ai}{\sqrt{z}} \: {\rm sign}(\dot h) \:  \right] 
			 \label{eq:Stokes-zeta-f} \,, \\
	\Xi_{\eta'} %& = \frac{\TR{\rho \sigma_2}}{\TR{\rho}}
			  & = - \frac{\alpha \Xi_\eta }{2 b \mathbb P }  \int \! \ud \tau \ud t \: \left[ \Ai_1 + \frac{2\Ai'}{z}  \right]  
			  \label{eq:Stokes-eta-f}\,,	\\
	\Xi_{\kappa'} %& = \frac{\TR{\rho \sigma_1}}{\TR{\rho}}
			  & = - \frac{\alpha \Xi_\kappa }{2 b \mathbb P } \int \! \ud \tau \ud t \:  \left[ (2g-1)  \Ai_1 +  g \, \frac{ 2 \Ai' }{z}  \right]  
			  \label{eq:Stokes-kappa-f}\,.
	\end{align}	
	The components of the final electron \Stokes vector along the electric field and wave vector in the rest frame, $\eta'$ and $\kappa'$,
	are proportional to their respective counterparts before the scattering. That means electrons which are initially unpolarized will not gain
	any degree of polarization along $\eta'$ or $\kappa'$. However, if the electrons are initially polarized along those directions their degree of
	polarization will change. 
	
	\subsubsection{Electron Polarization in an Ultrashort Pulse and Comparison with exact QED}	
	
	The \Stokes vector component
	along the magnetic field in the rest frame of the electron, $\Xi_\zeta'$, however, can become non-zero even for initially unpolarized
	electrons, $\Xi_\zeta=0$. The third term in the brackets in \eqref{eq:Stokes-zeta-f} contains the sign of the strong field shape function $\dot h$.
	Therefore, for a long monochromatic wave
	for each cycle of the wave the polarization builds up in the first half cycle and is reduced
	again in the second half cycle such that
	we expect a vanishing net-polarization in the end.
	However, we can expect some
	degree of polarization to survive for asymmetric pulses where these cancellations are not  complete. 
	Such an asymmetry might be realized, for instance, by using an ultra-short laser pulse.

	\begin{figure}[bt]
	\includegraphics[width=0.99\columnwidth]{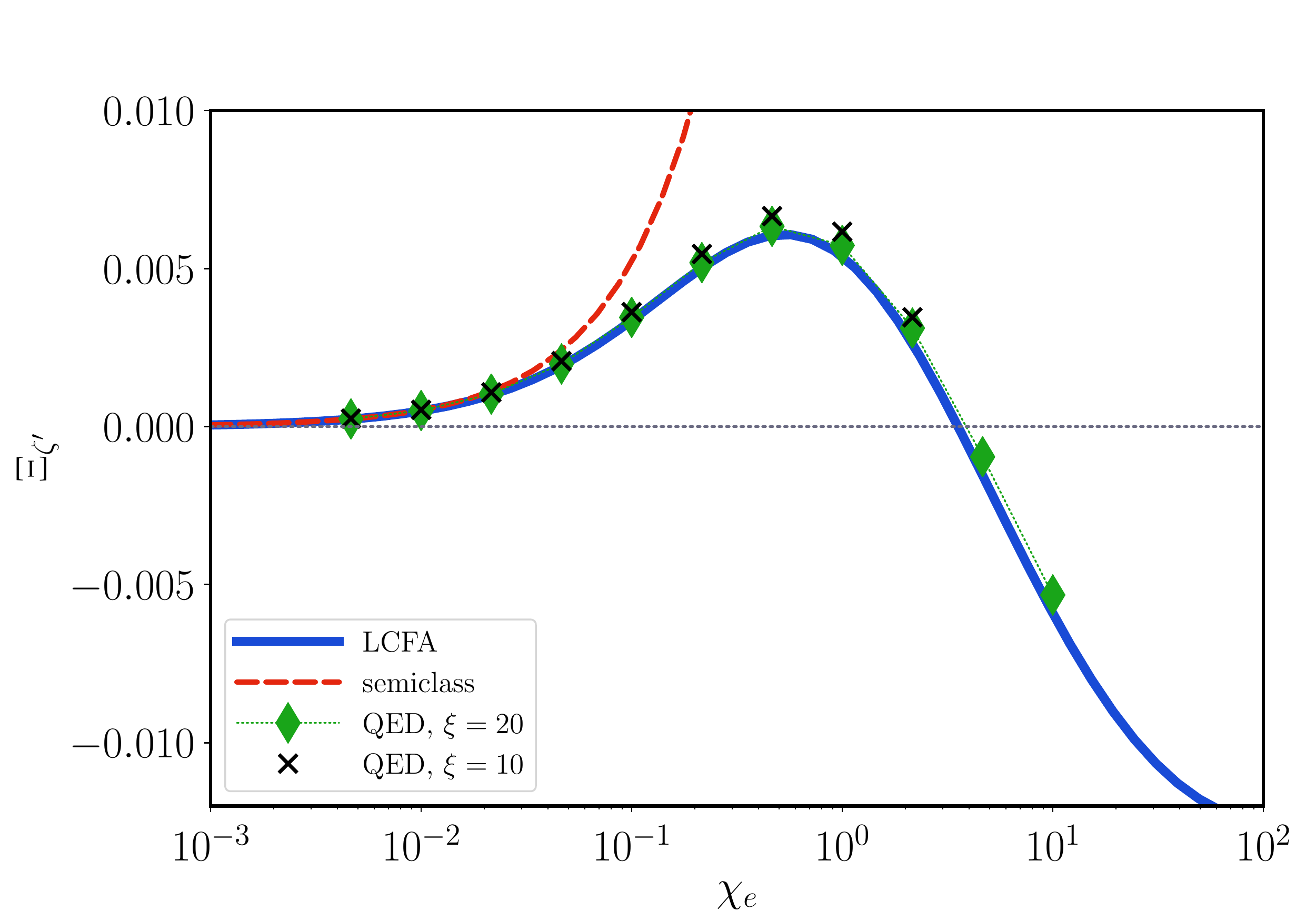}
	\caption{
	Final electron \Stokes parameter $\Xi_{\zeta'}$ as a function of $\chi_e$ for an ultrashort laser pulse with
	the optimal pulse duration $\Delta\phi = 3.27$.
	The blue solid curve is the LCFA result, the red dashed curve its semiclassical approximation \eqref{eq:Stokes-weakfield}.
	The green and black symbols represent numerical calculation of $\Xi_{\zeta'}$ in exact QED, using Eq.~\eqref{eq:rho-differential},
	without having done the LCFA approximation.
	}
	\label{fig:asymmetry}
	\end{figure}

	This becomes even more apparent if we expand the expression on the r.h.s.~of Eq.~\eqref{eq:Stokes-zeta-f} 
	in the semiclassical limit $\chi_e\ll 1$,
	\begin{align}
		\label{eq:Stokes-weakfield}
		\Xi_{\zeta'} &= - \frac{3 \chi_e}{10} \frac{\int \! \ud \tau \: |\dot h (\tau ) | \dot h (\tau ) }{\int \! \ud \tau \: |\dot h( \tau )| } 
		=  \frac{3 \chi_e}{10} \: \mathcal A \,,
	\end{align}
	i.e. the final electron polarization along the non-precessing direction is proportional to the quantum efficiency parameter
	and depends on the asymmetry of the pulse shape function. Let us write the pulse shape function (for the vector potential) as
	$h = \cos ( \phi + \phi_\mathrm{CE} )  \, \cos^2(\frac{\pi\phi}{2 \Delta\phi }) \Theta(  \Delta \phi - |\phi | )$.
	Note that we have to ensure $h(-\infty) = h(\infty)$, otherwise we are dealing with ``unipolar'' fields which (i)
	have been argued cannot be produced with lasers \cite{Milosevic:JPB2006} and (ii) cause infrared divergences in
	the scattering probabilities \cite{Ilderton:PRD2013}.
	This strongly limits the achievable asymmetries of the magnetic field, which has the shape function $\dot h$.
	For symmetry reasons the asymmetry of $\dot h$ vanishes for $\phi_\mathrm{CE} = 0$ and reaches its maximum
	for a carrier envelope phase of $\phi_\mathrm{CE} = \pi/2$. The maximum of the asymmetry
	of $|\mathcal A| \simeq 0.17$ is achieved for a pulse duration of $\Delta \phi \simeq 3.27$,
	and effectively becomes zero for $\Delta \phi > 8$.

	The blue solid curve in Fig.~\ref{fig:asymmetry} shows the value of the \Stokes parameter
	$\Xi_{\zeta'}$ as a function of $\chi_e$. For increasing $\chi_e$ the \Stokes parameter first increases and reaches a maximum
	at around $\chi_e = 0.5$. For even larger quantum parameters the \Stokes parameter gets smaller again and eventually switches sign for
	$\chi_e > 3.5$. 
    For comparison, the red dashed curve in Fig.~\ref{fig:asymmetry} is the weak field expansion for $\chi_e\ll1$, Eq.~\eqref{eq:Stokes-weakfield}.
    It shows that the semiclassical approximation,  which is linear in $\chi_e$,
    overestimates the true degree of polarization, and starts deviating significantly already for $\chi_e \lesssim 0.1$.
	In fact, already for $\chi_e\sim1$ the semiclassical result overestimates the full LCFA result about one order of magnitude.	
	
	In order to show the validity of the LCFA approximation for the polarization density matrix
	we also calculated $\Xi_{\zeta'}$ numerically from the exact QED polarization density matrix. 
	For the numerical integrations of Eq.~\eqref{eq:rho-differential} we used exactly the same pulse shape as for the LCFA calculation.
	These results are shown in Fig.~\ref{fig:asymmetry} 
	for $\xi=20$ as green diamond symbols and for $\xi=10$ as black crosses.
	The agreement between the the exact QED result and the LCFA is quite good already for the lower intensity, $\xi=10$, with a relative difference
	of $10\%$ at the peak, $\chi_e=0.5$.
	For $\xi=20$ the agreement is even better, with a relative difference of less than $5\%$.
	This behavior is to be expected since the LCFA formally coincides with exact QED the limit $\xi\to\infty$.
	A recent explicit comparison between exact QED and LCFA results applied to a Monte Carlo code confirmed similar trends and deviations
	for the energy loss of the electrons and photon emission angles \cite{Blackburn:2018}.

    There are some stark differences between the case studied here and the one investigated in  Sect.~\ref{sect:numerics}.
	In the latter calculation we found a relatively large degree of polarization in the scattering plane for scattering in a circularly polarized pulse, and
	this polarization persisted also for longer pulses. Here, in contrast, for linearly polarized pulses the polarization degree is much smaller and
	only occurs for ultra-short pulse durations.

	\subsubsection{The Spin-Flip and Non-Flip Rates}	
	\label{sect:flip-transitions}

	The probability rates for spin-flips are of special significance \cite{book:Sokolov,DelSorbo:PRA2017,DelSorbo:PPCF2018}.
	We have seen in the numerical results in Section \ref{sect:numerics}, Fig.~\ref{fig:on-axis2}, that
	the probabilities for spin-flip transitions strongly depend on the orientation of the initial electron spin vector.
	
	It is therefore not sufficient to know the spin-flip rates with regard to the non-precessing quantization
	direction $\vec \zeta$, instead we need to
	know the spin-flip and non-flip rates for an arbitrary initial polarization direction $\vec \Xi$ with, $|\vec \Xi| =1$.
	Having a non-flip transition means $\vec \Xi' = \vec \Xi$, while a spin-flip transition is characterized by $\vec \Xi' = - \vec \Xi$.
	By using the LCFA polarization density matrix, we evaluate the spin-flip and non-flip rates as
    \begin{widetext}
	\begin{align} 
	\mathbb P_\mathrm{non-flip} 
		&\equiv \tr \left[ \rho \frac{1+ \vec \Xi \cdot \vec \sigma}{2} \right]
		   = -\frac{\alpha}{ 2 b } \int \ud t \ud \tau \: \left( 
	2 \Ai_1 + (g+1)  \frac{2\Ai' }{z}  + \Xi_\zeta   {\rm sign}(\dot h)\:  \frac{2t-t^2}{1-t} \frac{\Ai }{\sqrt{z}}
		+ \Xi_\kappa^2 \frac{t^2}{1-t} \left[  \Ai_1 +  \frac{ \Ai' }{z}  \right] \right) \nonumber 
	\,, \\	   
	\mathbb P_\mathrm{flip}  \label{eq:LCFA-flip-nonflip}
		&\equiv \tr \left[ \rho \frac{1- \vec \Xi \cdot  \vec \sigma}{2} \right]
		   = -\frac{\alpha}{ 2 b } \int \ud t \ud \tau \:  
 		\frac{t^2}{1-t} \left( \left[  \frac{\Ai'}{z} 	- \Xi_\zeta   {\rm sign}(\dot h)\:    \frac{\Ai}{\sqrt{z}} \right]
 		- \Xi_\kappa^2 \left[  \Ai_1 +  \frac{ \Ai' }{z}  \right] \right)
           \,.
	\end{align}
    \end{widetext}
	These expressions depend linearly on $\Xi_\zeta$, and quadratically on $\Xi_\kappa$.
	That means the electrons will polarize
	only along the direction $\zeta$, i.e. the magnetic field in the rest frame of the particle. But the rates also depend on the
	degree of polarization along $\kappa$ (and implicitly along $\eta$).
	Any polarization along the $\kappa$ direction will decrease the spin-flip rates and increase the non-flip rates by the equal amount leaving
	the total rate unchanged. That means that the direction of the electron spin vector in relation to the strong field direction affects the spin-flip rates
	significantly. These above rates \eqref{eq:LCFA-flip-nonflip} could be implemented into spin-dependent
	Monte Carlo photon emission codes. They will allow to investigate the radiative polarization of electrons when
	multi-photon emissions in laser electron-beam collisions are important.

	\paragraph*{Semiclassical limit---}%
	Finally, we calculate the semiclassical limit
	of small quantum efficiency parameter $\chi_e \ll 1$, in order to connect our results to previous literature.
	We obtain for the spin-flip rate, up to terms $\mathcal O(\chi_e^3)$
	\begin{widetext}
	\begin{align}
	\mathbb R_\mathrm{non-flip}(\tau) &= 
	\frac{5}{2\sqrt{3}} \frac{\alpha\chi_e(\tau) }{b}
	\left(  1 - \chi_e(\tau) \left[ \frac{8}{5\sqrt{3}} + \frac{3}{10}  {\rm sign}(\dot h)\:  \Xi_\zeta\right] 
	+\chi_e^2(\tau) \left[ \frac{25}{8} + \frac{1}{12}\Xi_\kappa^2  + \frac{3\sqrt{3}}{5} 
	{\rm sign}(\dot h)\:  \Xi_\zeta \right] \right) \,,\\
	\mathbb R_\mathrm{flip}(\tau) &=  
	\frac{5}{2\sqrt{3}} \frac{\alpha \chi_e^3(\tau) }{b}
	\left( \frac{3}{8} - \frac{1}{12} \Xi_\kappa^2  + \frac{\sqrt{3}}{5} {\rm sign}(\dot h)\:  \Xi_\zeta \right)\,,
	\end{align}
	which agrees with a calculation of Baier and Katkov for the spin flip rates in
	a magnetic field for quantum parameter $\chi \ll 1$ \cite{Baier:JETP1967}. 	
	Here again, $\chi_e(\tau)$ refers to the local value of the quantum efficiency parameter.

	By adding the spin-flip and non-flip rates we find the total rate,
	\begin{align}
	\mathbb R = 
	\mathbb R_\mathrm{non-flip} + \mathbb R_\mathrm{flip} &= 
	\frac{5}{2\sqrt{3}} \frac{\alpha\chi_e(\tau) }{b}
	\left( 
		1 - \chi_e(\tau) \left[ \frac{8}{5\sqrt{3}} + \frac{3}{10} {\rm sign}(\dot h)\: \Xi_\zeta\right]  
		+ \chi_e^2(\tau) \left[ \frac{7}{2} + \frac{4\sqrt{3}}{5}  {\rm sign}(\dot h)\:  \Xi_\zeta \right]
	\right) \,.
	\end{align}
	\end{widetext}
 	By setting $\Xi_\zeta=0$ this expression agrees with textbook results for photon emission rate of unpolarized electrons,
 	see e.g.~Refs.~\cite{book:Baier,Ritus:JSLR1985}.

	\section{Summary \& Conclusion}
	\label{sect:summary}

	In this paper we investigated in detail the polarization properties of electrons in the non-linear Compton scattering process,
	when they are emitting a high-energy photon in an interaction with a high-intensity laser pulse. 

	Using the density matrix formalism allows to conveniently 
	keep track of the polarization properties of all involved particles, and for a unified treatment of unpolarized,
	partly polarized, and completely polarized electrons.	
	It is straightforward to calculate scattering probabilities and,
    the final electron \Stokes vector, or spin-flip rates from the density matrix.

	We investigated numerically the scattering of high-energy electrons from short intense laser pulses.
    In the case of circular laser polarization
	the resulting electron polarization is almost completely within the scattering plane, i.e.~the scattered electron beam will
	be radially polarized with polarization degrees up to $9\%$ after emitting a single photon for $\chi_e <10$.
	Since the electrons also lose energy during the photon emission
	this could be an accessible experimental signature for the polarization aspect of quantum radiation reaction
	in electron laser collisions.

	We derived the local constant crossed field approximation (LCFA) of the polarization density matrix
	as  a generalization of the known LCFA scattering rates. 
	These result will be useful to include spin-dependent QED effects into (PIC) laser-plasma simulation codes.
    The density matrix description of the quantum scattering process matches conceptually with the use of
    macro particles in PIC codes representing ensembles of electrons.
	With those codes, in turn, multi-photon emission effects can be studied for radiative spin-polarization
    in high-intensity laser-matter interactions
    such as the electron beam conditioning in laser plasma accelerators.
    In addition, the impact of electron spin polarization on radiation reaction, QED cascade formation, and
   	in the strong fields around magnetars could be investigated.

    We have explicitly verified that the value of the \Stokes vector calculated using the LCFA approximation of the
    polarization density matrix agrees well with an exact QED calculation for $\xi \gtrsim 10$. Moreover, 
    the semiclassical approximation of the LCFA expressions, $\chi_e \ll 1$,
    commonly used to calculate the equilibrium spin polarization of electrons in storage rings \cite{Mane:RPP2005},
    deviate from the full LCFA and the
    exact QED results already for $\chi \lesssim 0.1$ and is therefore inadequate for making predictions for
    the spin-polarization in future high-intensity laser experiments aiming to reach $\chi_e\sim 1$.

    The calculations of the spin-dependent non-linear Compton scattering in this paper are based on Volkov electrons,
    which contain the interaction with the background laser field to all orders, but do not contain radiative corrections.
    Therefore Volkov electrons do not possess any anomalous magnetic moment $a_e = ( g_e - 2 )/2 = 0$.
    There are, however, small contributions to the spin-dependent scattering rates
    due to the anomalous magnetic moment $\propto a_e = \mathcal O(10^{-3})$,
    which are not considered here \cite{Mane:RPP2005}. These corrections might become important for other
    particles with larger anomalies. %, e.g.~protons.
    We also note that the anomalous moment itself becomes field dependent $a_e(\chi_e)$
    due to the field-dependence of the electron self energy \cite{Ritus:AnnPhys1972}.
    Detailed investigations of these small corrections to our results are left for future work.

	\begin{acknowledgements}

	DS acknowledges valuable discussions with Jonathan Gratus, Tom Heinzl, Anton Ilderton,
    Ben King, and Maxim Korostelev, and 
	support from the Science and Technology Facilities Council, Grant No. ST/G008248/1. 
	CPR and DDS acknowledge support from Engineering and Physical Sciences grant EP/M018156/1.
	AGRT acknowledges support from U.S. DOD under Grant No. W911NF-16-1-0044.
	\end{acknowledgements}

	\appendix

	\section{The Traces over the Dirac Structures}
	\label{sect:traces}	
	
	In this appendix we give some details on calculating the traces over the Dirac structures, and collect all relevant results.
	The Dirac algebra of the gamma matrices is defined by their anticommutator $\{ \gamma^\mu , \gamma^\nu \} = 2 g^{\mu\nu} $.
	In addition, we need, for the evaluation of the traces involving the spin four-vectors, the matrix 
	$\gamma^5 = - \frac{i}{4!} \epsilon_{\mu\nu\alpha\beta} \gamma^\mu \gamma^\nu \gamma^\alpha \gamma^\beta$,
	which anti-commutes with all $\gamma^\mu$, $\{ \gamma^5 , \gamma^\mu \} = 0$,
	is hermitian $(\gamma^5)^\dagger = \gamma^5$, and fulfills $\gamma^5 \gamma^5 = 1$.
	The completely antisymmetric Levi-Civita tensor $\epsilon^{\mu\nu\alpha\beta}$ has components
	$\epsilon^{+-12} = -2$ in light-front coordinates.
	The Dirac adjoint $\bar \Gamma$ for any element $\Gamma$ of the Dirac algebra
	is given by $\bar \Gamma = \gamma^0 \Gamma^\dagger \gamma^0$.

	The final electron polarization density matrix Eq.~\eqref{eq:final-density-matrix-allintegrals} depends on only the following
    four independent traces over the Dirac matrices. Expressions for these traces can be found in textbooks on
    quantum field theory, e.g.~\cite{book:Itzykson}. We only quote the traces involving $\gamma^5$, which are less common.
	Only those traces with an even number of four or more $\gamma^\mu$ matrices in addition to $\gamma^5$ are nonzero.
    The relevant ones read
    \begin{widetext}
	\begin{align}
	\tr_\mathrm{D} \big[ \gamma^\mu \gamma^\nu \gamma^\alpha \gamma^\beta \gamma^5 \big]
		&= -4i \epsilon^{\mu\nu\alpha\beta} \,, \\
	\tr_\mathrm{D} \big[ \gamma^\mu \gamma^\nu \gamma^\alpha \gamma^\beta \gamma^\sigma \gamma^\tau \gamma^5 \big]
		&= - 4i \big[
		 g^{\mu\nu} \epsilon^{\alpha\beta\sigma\tau} 
		 - g^{\mu\alpha} \epsilon^{\nu\beta\sigma\tau} 
		 + g^{\nu\alpha} \epsilon^{\mu\beta\sigma\tau}
		 +g^{\beta\sigma} \epsilon^{\mu\nu\alpha\tau}
		 - g^{\beta\tau} \epsilon^{\mu\nu\alpha\sigma} 
		 +g^{\sigma\tau} \epsilon^{\mu\nu\alpha\beta}
		\big] \,. \label{eq:trace-6}
	\end{align}

    	By using $\mathscr J^\mu$ given in Eq.~\eqref{eq:current-integrand} the relevant Dirac traces are given by
		\begin{align} 
	\mathsf{UP} & \equiv
		 \frac{1}{4} \tr_\mathrm{D} \big[ ( \slashed p' + m) \: \mathscr J^\mu(\phi) \:
		 							 (\slashed p+m) \: \bar{ \mathscr J }_\mu(\phi') \big]
		  =  2m^2 + m^2\xi^2 g(u) [ h(\phi') - h(\phi) ]^2  \,,
		   \label{eq:UP-trace} \\
		\mathsf{IP}(S)  & \equiv \nonumber
		  \frac{1}{4} \tr_\mathrm{D} \big[  ( \slashed p' + m) \: \mathscr J^\mu(\phi) \:
		  							 (\slashed p+m) \: \gamma^5\slashed S \: \bar{ \mathscr J }_\mu(\phi') \big] \\
			&= 
			\frac{2im^2\xi}{(k.p)} \: [h(\phi') - h(\phi) ] 
			\left\{
			 (p'.\tilde f .S) \left[ 1 + \frac{u}{2}\right] - (p.\tilde f. S)
			\right\} \,,	 
			\label{eq:IP-trace} \\
	\mathsf{FP}(S') & \equiv \nonumber
		 \frac{1}{4} \tr_\mathrm{D} \big[ ( \slashed p' + m) \: \gamma^5\slashed S' \: \mathscr J^\mu(\phi)  \:
		 							 (\slashed p+m) \: \bar{ \mathscr J }_\mu(\phi') \big]  \\
		 & = 
		 	 \frac{2 i m^2 \xi}{(k.p)} \: [h(\phi') - h(\phi)] 
				\left\{
				  (p'.\tilde f. S') \, [1+u] - (p.\tilde f .S') \left[ 1 + \frac{u}{2} \right]
				\right\} \,,
		 \label{eq:FP-trace}  \\
	 \mathsf{PC}(S' , S) & \equiv \nonumber 
	      \frac{1}{4} \tr_\mathrm{D} \big[  ( \slashed p' + m )\: \gamma^5\slashed S' \: \mathscr J^\mu(\phi) \: 
	      								(\slashed p+m)  \: \gamma^5\slashed S \: \bar{ \mathscr J }_\mu(\phi') \big] \\		
	      	& = 	  -2 m^2 (S.S') 
								 + m^4 \xi^2 \frac{(S.k)(S'.k)}{(k.p)^2} 
								 \left\{  	(1+u) [h(\phi')-h(\phi)]^2 - u^2 \, h(\phi')h(\phi) 	 \right\} \nonumber  \\
							&  \quad	-  m^3 \xi \, u \: [ h(\phi') + h(\phi)  ] \: \frac{(S.f.S')}{(k.p)} 
							 -  m^2 \xi^2   \frac{1+u}{(k.p)^2} [h(\phi') - h(\phi)]^2 \nonumber \\
							&  \quad \times
							     		\left\{
											(p.p')(S.k)(S'.k) + (S.S')(p.k)(p'.k)
												- (p.S') (p'.k)(S.k)  - (S.p') (p.k) (S'.k)
										\right\}  	\label{eq:PC-trace}		\,,							
	\end{align}
    \end{widetext}
	where $\mathsf{UP}$ stands for \textit{unpolarized}, because the corresponding expression is independent of the
	polarization of the electrons. On the contrary, $\mathsf{IP}$ ($\mathsf{FP}$) refers to trace expressions
	which depend on the \textit{initial (final) electron polarization}. The corresponding expressions
	depend on the space-like spin-vector $S^\mu$ and $S'^\mu$, with $S.S=-1$, respectively. We note here that
	we could choose \textit{any} valid space-like unit vector that fulfils $S.p=0$ ($S'.p'=0$).
	However, evaluating the scalar products in $\mathsf{IP/FP}$ becomes particularly
	simple when we use the physical basis introduced in Section \ref{sect:basis}, for details of these calculations see Appendix \ref{sect:transverse:integral}.
	Finally, the term $\mathsf{PC}$ describes the \textit{polarization correlation} between the initial and final electrons.
	We note that additional traces need to be considered when taking into account also the polarization of the
	{emitted photons} as well.

	Additional definitions used in the above expressions are $u = (k.k')/(k.p') = t/(1-t)$, $t = (k.k')/(k.p) $ and
	\begin{align} \label{eq:def-g}
	g = 1 + \frac{u^2}{2(1+u)} =1 + \frac{t^2}{2(1-t)} \,.
	\end{align}
	Moreover, $h$ is the shape function of the linearly polarized laser vector potential $A^\mu = m \xi \varepsilon^\mu h(\phi)$
	with $\varepsilon^\mu$ being the polarization vector,
	$f^{\mu\nu} =k^\mu\varepsilon^\nu - k^\nu \varepsilon^\mu$,
	$\tilde f^{\mu\nu} = \frac{1}{2} \epsilon^{\mu\nu\alpha\beta} f_{\alpha\beta}$,
	and $(S.f.S') \equiv S_\alpha f^{\alpha\beta} S'_\beta$.

	\section{Details on the Transverse Momentum Integrals}
	\label{sect:transverse:integral}

	Here we collect relations needed for the Gaussian integrals over the transverse photon momentum
	of the polarization density matrix, Eq.~\eqref{eq:final-density-matrix-allintegrals}.
	In \cite{Dinu:PRA2013} these integrals were explicitly performed to derive analytic
	expressions for the unpolarized total nonlinear Compton probability.
	Here, these results are generalized by including also the electron spin polarization.
	
	\subsection{The Exponential Part}
	
\begin{figure}[bt]
	\includegraphics[width=0.99\columnwidth]{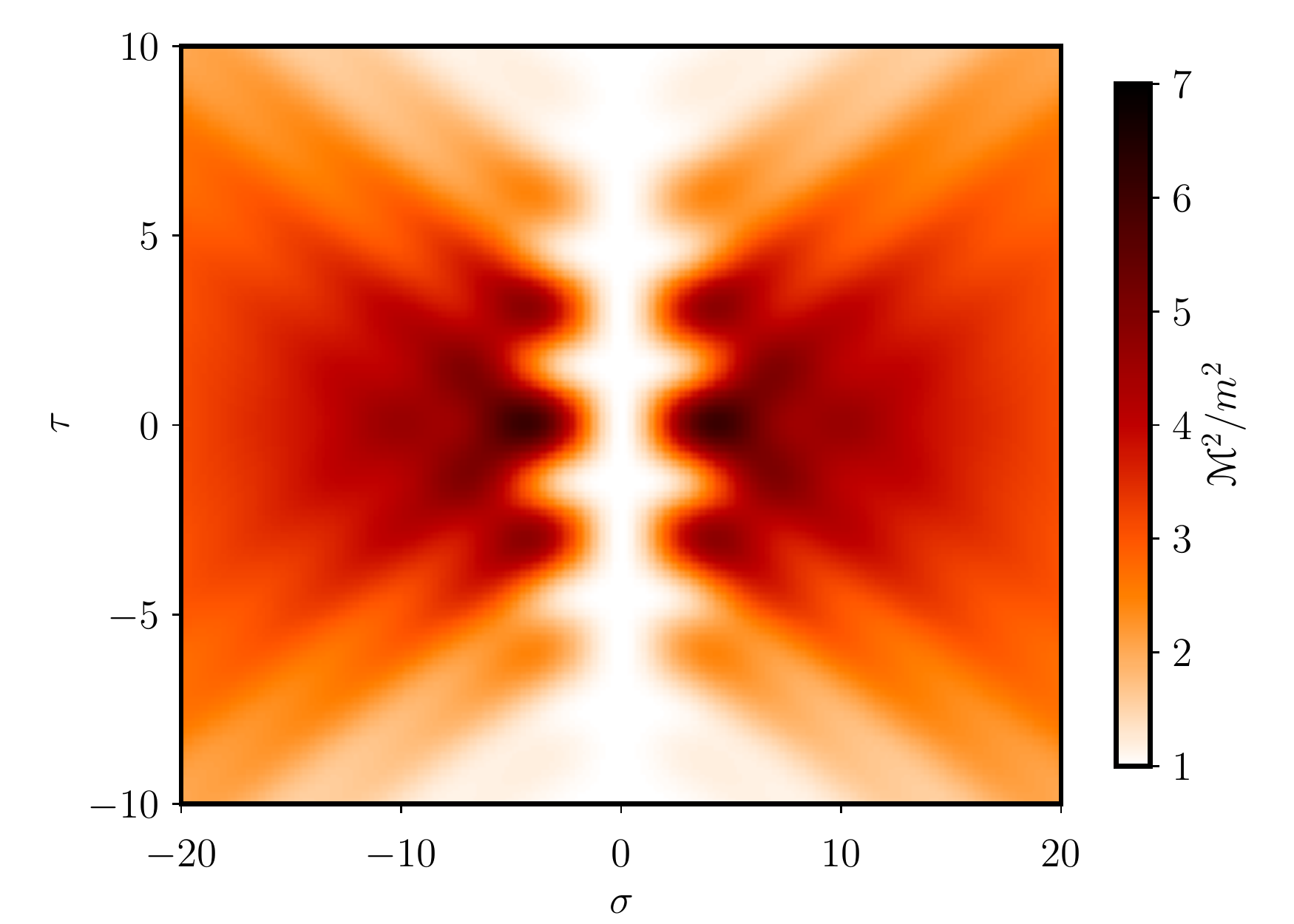}
	\caption{Kibble's mass $\mathscr M^2$ for $\xi = 3$. It approaches $\mathscr M^2\to m^2$ for small values of $\sigma$.}
	\label{fig:kibblemass}
	\end{figure}

	First, we note that the term in the exponent of \eqref{eq:final-density-matrix-allintegrals} can be rewritten as follows:
	\begin{multline} \label{eq:exponent-Kibble}
	\frac{k' . \langle\pi\rangle}{k.p'} 
						=
					\frac{ \xi t  }{2  m^2 \chi_e (1-t)}  \bigg( \mathscr M^2  \\
						+ \frac{[\vec k_\perp'\cdot \vec\varepsilon - t\langle \vec \pi_\perp\cdot \vec\varepsilon\rangle ]^2}{ t^2 } 
						+ \frac{[\vec k_\perp'\cdot \vec\beta         - t\langle \vec \pi_\perp\cdot \vec\beta\rangle ]^2}{ t^2 } 
						\bigg)
	\end{multline}
	with Kibble's effective mass \cite{Kibble:NPB1975,Harvey:PRL2012,DiPiazza:PRD2018} that appears in 
	the gauge invariant part of the Volkov propagator and in the Wigner function \cite{Hebenstreit:PRD2011a}:
	\begin{align} \label{eq:kibble:mass}
	\mathscr M^2   & =m^2 + \langle \vec \pi_\perp^2\rangle - \langle \vec \pi_\perp \rangle^2 
     						 = m^2 (1 + \xi^2 \langle h^2 \rangle -  \xi^2 \langle h \rangle^2 ) 
     						 \,.
	\end{align}
	The Kibble mass is a function of the variance of the laser vector potential
    (with regard to the \floating average over an phase interval of length $\sigma$ around $\tau$)
	and depends on both $\sigma$ and $\tau$. Its dependence on those variables is plotted in Fig.~\ref{fig:kibblemass}
	for $\xi=3$ for a pulse shape of $h = \sin \phi \, \cos^2(\frac{\pi\phi}{8\pi}) \Theta(   4\pi - |\phi | )$.
	We see in Fig.~\ref{fig:kibblemass} that  $\mathscr M^2 \to m^2$ for $\sigma\to 0$.
	Only for an infinitely long monochromatic wave, $h=1$, the Kibble mass approaches
	the usual intensity-dependent effective mass $\mathscr M^2 \to m_\star^2 = m^2 (1 + \xi^2/2)$ for $\sigma \to \infty$,
	and it goes to zero for finite pulses \cite{Harvey:PRL2012}.
	
	By using Eq.~\eqref{eq:exponent-Kibble}, we see that the transverse momentum integral over $\ud^2 \vec k_\perp' $
	appearing in \eqref{eq:final-density-matrix-allintegrals} is 
	Gaussian, with the value
	\begin{align}
	\mathcal G_0 
    	& \equiv	\int_{\mathbb R^2} \! \d^2 \vec k'_\perp \: e^{i\sigma \frac{k'. \langle\pi\rangle}{k.p'}} \nonumber \\
	 	& = \frac{2\pi i }{\sigma +i0^+}  \frac{m^2 \chi_e}{\xi }\: t(1-t) \:  
	 	e^{ i \sigma \frac{ \xi  \mathscr M^2  }{2 \chi_e m^2} \frac{t}{1-t} }
	 \label{eq:G-0} \,,
	\end{align}
	where the infinitesimal imaginary part added to $\sigma$ ensures the convergence of the integral. 
	For unpolarized electrons this is the only structure that appears in the scattering probability \cite{Dinu:PRA2013},
	which gives we immediately that
	\begin{align}
	\int \! \ud^2 \vec k_\perp' e^{i\sigma \frac{k'.\langle \pi\rangle}{k.p'}} \: \mathsf{UP} = \mathcal G_0 \, \mathsf{UP} \,,
	\end{align}
	because $\mathsf{UP}$ does not depend on $\vec k_\perp'$.

	However, when we take into account the electron polarization additional terms occur, in which 
	$k'.\beta         = -  \vec k'_\perp \cdot \vec \beta_\perp $ and 
	$k'.\varepsilon = -  \vec k'_\perp \cdot \vec \varepsilon_\perp $
	stand in the pre-exponential under the integral. The corresponding transverse momentum integrals
	\begin{align}	
	\mathcal G_\varepsilon & \equiv	\int \! \d^2 \vec k'_\perp \: (k'.\varepsilon) \: e^{i\sigma \frac{k'. \langle\pi\rangle}{k.p'}}
	= t \, [ ( p .\varepsilon ) + m \xi \langle h \rangle ] \: \mathcal G_0
	\label{eq:G-epsilon} \,,	\\
	\mathcal G_\beta & \equiv 	\int \! \d^2 \vec k'_\perp \: (k'.\beta) \: e^{i\sigma \frac{k'. \langle\pi\rangle}{k.p'}}
	=  t (p.\beta) \: \mathcal G_0 
	\label{eq:G-beta} \,,
	\end{align}
	are proportional to $\mathcal G_0$.

	\subsection{Initial Polarized}
	
	The initial polarized traces appear only in the diagonal elements of the
    polarization density matrix \eqref{eq:final-density-matrix-allintegrals}.
	At first we evaluate the $\mathsf{IP}(s)$ trace, Eq.~\eqref{eq:IP-trace},
    for the three basis vectors $\{\zeta,\eta,\kappa\}$ introduced in Section
	\ref{sect:basis}. We obtain
	\begin{align}
	\label{eq:IP-zeta}
	\mathsf{IP}(\zeta) &= 	 im^2\xi \: t \: \sigma \langle \dot h\rangle \,, \\
	\label{eq:IP-eta}
	\mathsf{IP}(\eta) &= 0 \,, \\
	\label{eq:IP-kappa}
	\mathsf{IP}(\kappa) &=  im^2 \xi  \:  \sigma \langle \dot h \rangle
							\frac{2-t}{1 - t}
							\left\{ 
							 t (p\beta) -  (k'\beta)
							\right\}  \,.
	\end{align}
	Thus, by expanding the initial electron \Stokes vector in the physical spin basis
	we find $\mathsf{IP}(\Xi) = \Xi_\zeta \mathsf{IP}(\zeta) + \Xi_\kappa \mathsf{IP}(\kappa)$.

	Upon performing the transverse momentum integral
	\begin{align}
	\int \! \ud^2 \vec k_\perp' e^{i\sigma \frac{k'.\langle \pi\rangle}{k.p'}} \: \mathsf{IP}(\Xi) 
    =   im^2\xi \: t \: \sigma \langle \dot h\rangle \: \mathcal G_0 \: \Xi_\zeta \,,
	\end{align}
	where we used that
	\begin{align}
	\int \! \ud^2 \vec k_\perp' e^{i\sigma \frac{k'.\langle \pi\rangle}{k.p'}} \: \mathsf{IP}(\kappa)   = 0 \,,
	\end{align}
    because of the relation between \eqref{eq:G-0} and \eqref{eq:G-beta}.

	\subsection{Final Polarized}
	
	The final polarized traces appear in the diagonal ($\mathsf{FP}(\zeta')$) and off-diagonal ($\mathsf{FP}(\kappa')$,
    $\mathsf{FP}(\eta')$)
	elements of the final electron polarization density matrix, with
	\begin{align}
	 \mathsf{FP}(\zeta')  & =  im^2  \xi \: u \: \sigma  \langle \dot h \rangle \,, \\
	 \mathsf{FP}(\eta')  &= 0 \,, \\
	 \mathsf{FP}(\kappa') &=
						 im^2\xi \: \sigma \langle \dot h \rangle  \frac{2-t}{1-t}
							\left\{ 
							 t (p\beta) -  (k'\beta)
							\right\} \,.
	\end{align}

	The diagonal term containing $\mathsf{FP}(\zeta')$ can be easily integrated because it is independent of $\vec k_\perp'$,
	yielding
	\begin{align}
	\int \! \ud^2 \vec k_\perp' e^{i\sigma \frac{k'.\langle \pi\rangle}{k.p'}} \: \mathsf{FP}(\zeta')
	&= \mathsf{FP}(\zeta') \: \mathcal G_0
	  =  im^2  \xi \: u  \: \sigma \langle \dot h \rangle \: \mathcal G_0 \,,
	\end{align}		
	and the off-diagonal terms integrate to zero,
	\begin{align}
	\int \! \ud^2 \vec k_\perp' e^{i\sigma \frac{k'.\langle \pi\rangle}{k.p'}} \: \mathsf{FP}(\kappa')    = 0 \,,
	\end{align}	
	due to the relation between \eqref{eq:G-0} and \eqref{eq:G-beta}.

    \subsection{Polarization Correlation}

	\paragraph*{Diagonal Terms---}%
	By evaluating the expressions in \eqref{eq:PC-trace} for $s'=\zeta'$ and $s \in \{\zeta,\eta,\kappa\}$ we easily establish
	that $\mathsf{PC}(\zeta',\eta)=0$, and therefore
	\begin{align}
	\mathsf{PC}(\zeta',\Xi)  =  \Xi_\zeta \, \mathsf{PC}(\zeta',\zeta) + \Xi_\kappa \, \mathsf{PC}(\zeta',\kappa) \,,
	\end{align}
	with 
	\begin{align}
	\mathsf{PC}(\zeta',\zeta) &= 2m^2 + m^2 \xi^2 \sigma^2 \langle \dot h\rangle^2 \,, \\
	\mathsf{PC}(\zeta',\kappa) &= -2m^2 (1+u) \left\{ t (p.\beta) - (k'.\beta) \right\} \,.
	\end{align}
	
	By using the same arguments as above we obtain that
	\begin{align}
	\int \! \ud^2 \vec k_\perp' e^{i\sigma \frac{k'.\langle \pi\rangle}{k.p'}} \: \mathsf{PC}(\zeta',\Xi)
	& =  \mathcal G_0 \: \Xi_\zeta \, \mathsf{PC}(\zeta',\zeta) \,,
	\end{align}		
	i.e.~only the components of the initial electron \Stokes vector along $\zeta$ (i.e.~along the magnetic field in the
	rest frame of the initial electron) contribute to the diagonal elements of the polarization density matrix.

	\paragraph*{Off-Diagonal Terms---}%    	
	In the lower left element we have the trace (the upper right element is just the complex conjugate of that)
	\begin{multline}
	\mathsf{PC}(\kappa'+i\eta' , \Xi)  =   
									\Xi_\zeta \mathsf{PC} (\kappa' ,\zeta) 
                                   +  \Xi_\eta \mathsf{PC} (\kappa'+i\eta',\eta) \\
                                   + \Xi_\kappa \mathsf{PC} (\kappa'+i\eta' ,\kappa)
	\end{multline}
	with	
	\begin{align*}
	\mathsf{PC}(\eta',\zeta)     &= 0 \,,\\
	\mathsf{PC}(\kappa',\zeta) &= 2m^2 \left\{ t (p.\beta) - (k'.\beta) \right\} 
		\stackrel{\int\! \ud^2 \vec k'_\perp}{\longrightarrow} 0 \,,\\
	\mathsf{PC}(\eta',\eta) 
			&= 2m^2   + m^2 \xi^2 \sigma^2 \langle \dot h \rangle^2 \,,\\
	\mathsf{PC}(\kappa',\eta) 
			&
			   = 2m \left\{ \frac{(p.f.p')}{(k.p)}  + \frac{ m \xi}{2} \, t \: [h(\phi')+h(\phi)] \right\} \\
	\mathsf{PC}(\eta',\kappa) &= 
		 - 2m (1+u) \left\{ \frac{(p.f.p')}{(k.p)} \right. \\
               & \qquad \qquad \left. + \frac{ m \xi}{2}\, t \: [h(\phi')+h(\phi)] \right\} \\
	\mathsf{PC}(\kappa',\kappa) &=   4 m^2 g
	-2m^2 +  m^2\xi^2 g \, \sigma^2 \langle \dot h \rangle^2 \,.
	\end{align*}
	The two traces $\mathsf{PC}(\eta',\eta)$ and $\mathsf{PC}(\kappa',\kappa)$ are independent of $\vec k_\perp'$
	and their transverse momentum integral reduces to the multiplicative factor $\mathcal G_0$.
	Now we can evaluate the non-trivial transverse momentum integrals,
    \begin{align}
	\int \! \ud^2 \vec k_\perp'  e^{i\sigma \frac{k'.\langle \pi\rangle}{k.p'}} \mathsf{PC}(\eta',\kappa)
	&	=   2m^2\xi \frac{t}{1-t} \mathcal G_0 \, \delta h  \,, \\
	\int \! \ud^2 \vec k_\perp'  e^{i\sigma \frac{k'.\langle \pi\rangle}{k.p'}} \mathsf{PC}(\kappa',\eta)
	& = - 2m^2\xi t \mathcal G_0 \, \delta h \,,
	\end{align}
 	which do not vanish identically and contain new structures. They are
	proportional to $\delta h  = \langle h\rangle - \bar h$,
	the difference of the
	\floating average $\langle h \rangle$
	and the arithmetic average $\bar h =  \frac{ h(\phi')+h(\phi) }{2}$
	of the laser vector potential shape function $h$ between the two
	phase points $\phi$ and $\phi'$. We can therefore assume that this difference vanishes when the laser field
	can be considered as constant. But in general this could be a source of physics beyond the constant crossed
	field approximation.

\begin{figure}[!thb]
	\includegraphics[width=\columnwidth]{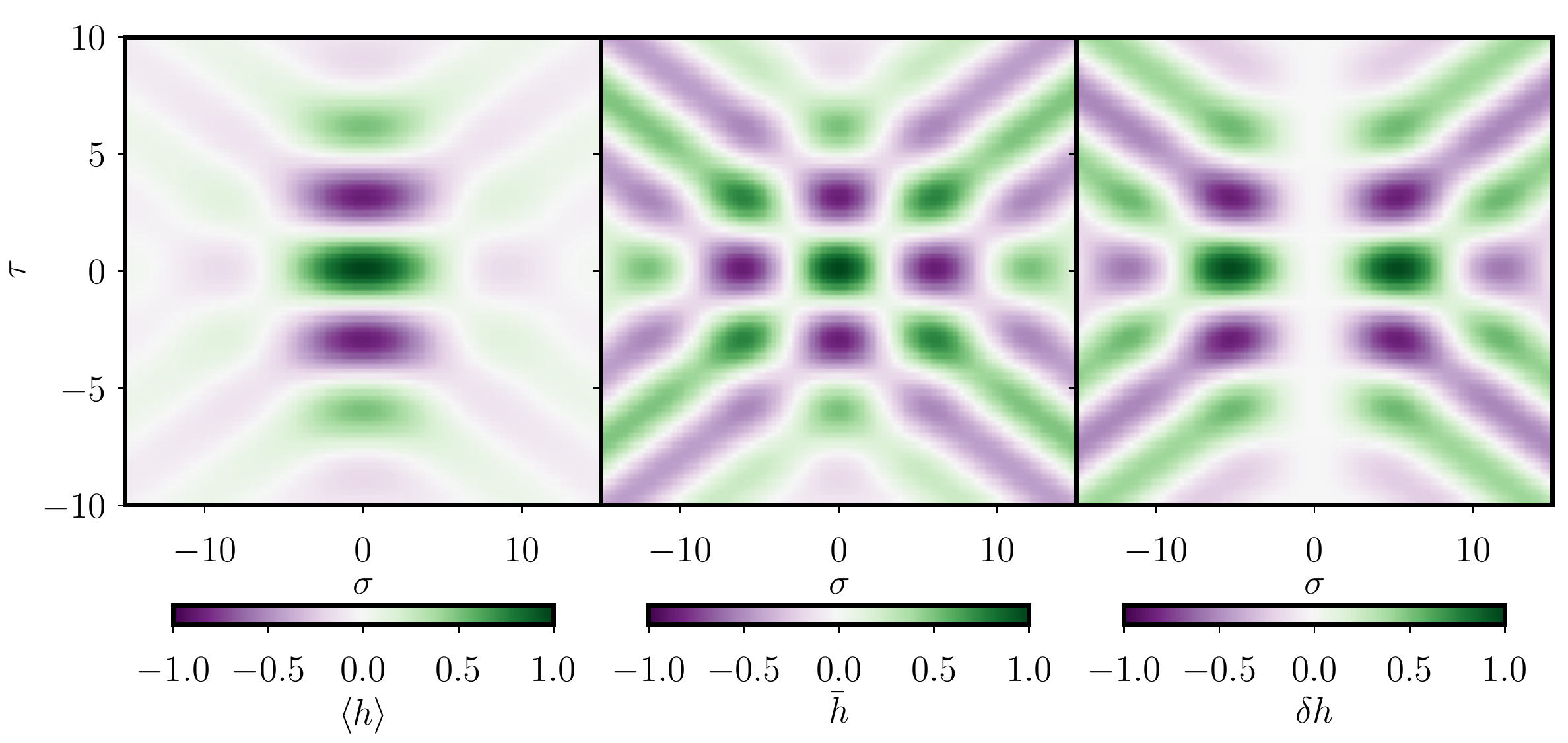}	
	\caption{The \floating average $\langle h \rangle$, the arithmetic average $\bar h$, and their difference $\delta h$.}
	\label{fig:averages}
	\end{figure}

	Lets make this a bit more explicit:
	Expanding $\delta h$ in powers of $\sigma = \phi'-\phi $ around the midpoint $\tau = (\phi'+\phi)/2$
	yields, to lowest order, $\delta h   \simeq - \frac{\sigma^2}{12} \ddot h(\tau)$.
	This shows that $ \delta h  = 0$ for a constant crossed field, because $\dot h=const.$, and $\ddot h$ and all higher
	derivatives vanish.
	The various averages are plotted in Fig.~\ref{fig:averages} as a function of $\sigma$ and $\tau$, showing that indeed $\delta h \approx 0$ for
	 $\sigma \ll 1$.

	Putting all the results from this Appendix together yields Eq.~\eqref{eq:final-density-matrix-kperp-done} for the transverse momentum integrated 
   final electron polarization density matrix.

%\bibliography{references}

\end{document}